\documentclass[aps,prl,showpacs,superscriptaddress,twocolumn]{revtex4}

\usepackage{blindtext}
\usepackage[utf8]{inputenc}
 
 \usepackage{hyperref}
 \usepackage{bbm}
 \usepackage{graphicx}
 \usepackage{amsmath}
 \usepackage{xy}
 \usepackage{amssymb}
 \usepackage{mathscinet}
 \usepackage{amsfonts,amsmath}
 \usepackage{xparse}
 \usepackage{bbold}

\def\ket#1{| #1 \rangle}

\newcommand{\sbigotimes}{%
  \mathop{\mathchoice{\textstyle\bigotimes}{\bigotimes}{\bigotimes}{\bigotimes}}%
}

\begin{document}

\title{Heat-Bath Algorithmic Cooling  with correlated qubit-environment interactions}

\author{Nayeli A. Rodr\'iguez-Briones}
\email{narodrig@uwaterloo.ca}
\affiliation{Institute for Quantum Computing, University of Waterloo, Waterloo, Ontario, N2L 3G1, Canada}
\affiliation{Department of Physics \& Astronomy, University of Waterloo, Waterloo, Ontario, N2L 3G1, Canada}
 
 \author{Jun Li}
 \affiliation{Beijing Computational Science Research Center, Beijing 100193, China}


\author{Xinhua Peng}
\affiliation{Hefei National Laboratory for Physical Sciences at Microscale and Department of Modern Physics, University of Science and Technology of China, Hefei, Anhui 230026, China}
\affiliation{Synergetic Innovation Center of Quantum Information \& Quantum Physics,
University of Science and Technology of China, Hefei, Anhui 230026, China}

\author{Tal Mor}
\affiliation{Computer Science Department, Technion, Haifa 320008, Israel}

\author{Yossi Weinstein}
\affiliation{Computer Science Department, Technion, Haifa 320008, Israel}

\author{Raymond Laflamme}
\email{laflamme@iqc.ca}
\affiliation{Institute for Quantum Computing, University of Waterloo, Waterloo, Ontario, N2L 3G1, Canada}
\affiliation{Department of Physics \& Astronomy, University of Waterloo, Waterloo, Ontario, N2L 3G1, Canada}
\affiliation{Perimeter Institute for Theoretical Physics, 31 Caroline Street North, Waterloo, Ontario, N2L 2Y5, Canada}
\affiliation{Canadian Institute for Advanced Research, Toronto, Ontario M5G 1Z8, Canada}

\date{\today}

\begin{abstract}
Controlled preparation of highly pure quantum states is at the core of practical applications of quantum information science, from the state initialization of most quantum algorithms to a reliable supply of ancilla qubits that satisfy the fault-tolerance threshold for quantum error correction.
Heat-bath algorithmic cooling has been shown to purify qubits by controlled redistribution of entropy and multiple contact with a bath, not only for ensemble implementations but also for technologies with strong but imperfect measurements. However, an implicit restriction about the interaction with the bath has been assumed in previous work.
In this paper, we show that better purification can be achieved by removing that restriction.
More concretely, we include correlations between the system and the bath, and we take advantage of these correlations to pump entropy out of the system into the bath.
We introduce a tool for cooling algorithms, which we call ``state-reset'', obtained when the coupling to the environment is generalized from individual-qubits relaxation to correlated-qubit relaxation.
We present improved cooling algorithms which lead to an increase of purity beyond all the previous work, and relate our results to the Nuclear Overhauser Effect.  
\end{abstract}

\maketitle



\section{I. Introduction}
Quantum information processing brings ways for cooling physical systems by
manipulating entropy in an algorithm way~\cite{schulman1999molecular,boykin2002algorithmic,fernandez2004algorithmic,schulman2005physical,schulman2007physical}. 
Understanding these processes and their cooling limits can elucidate fundamental theoretical properties of quantum thermodynamics and lead to new experimental possibilities. In particular, \textit{algorithmic cooling} 
methods have important applications in quantum computing 
as they provide a potential solution to prepare quantum systems with sufficient purity. Controlled preparation of highly pure quantum states is essential for a reliable supply of ancilla qubits in quantum error correction, and is at the core of many
quantum algorithms on the initialization phase.
While algorithmic cooling has mainly focused on
ensemble quantum computing implementations~\cite{chang2001nmr,fernandez2005paramagnetic,elias2011heat,brassard2014prospects,baugh2005experimental,Ryan:2008qf,park2014hyperfine,park2015heat}, 
it could also be used to increase the purity of initial states
up to the fault-tolerance threshold for technologies with strong but imperfect projective measurements (e.g. in superconducting qubits).
This technique could be also used to complement randomized benchmarking techniques by
distinguishing state and measurement errors.
Another potential use of algorithmic cooling is for improving
signal to noise ratio in NMR and MRI
applications~\cite{fernandez2005paramagnetic,mor2005patent}
(see also their limitations analyzed in~\cite{brassard2014prospects}).

S{\o}rensen~\cite{sorensen1991entropy} was the first to observe the constraint of unitary dynamics to increase the polarization $\epsilon=\operatorname{Tr}[\rho Z]$ ($\epsilon$ is related to purity $\mathcal{P}$ by $\epsilon=\sqrt{2\mathcal{P}-1}$ in the basis that $\rho$ is diagonal), for the density matrix $\rho$ and the Pauli operator $Z$, of a subset of qubits at the expense of decreasing the polarization of the complementary qubits. 
In the context of quantum information, Schulman and Vazirani proposed cooling
algorithms and coined the term ``quantum mechanical heat
engine"~\cite{schulman1999molecular}, which was inspired by Peres's recursive algorithm~\cite{peres1992_590} of von Neumann's extraction of fair coin flips from a sequence of biased ones~\cite{von195113}. 
This heat engine carries out a reversible entropy compression process in which an input of energy to the system results in a separation of cold and hot regions.  
Furthermore, an explicit way to implement entropy compression in ensemble quantum computers (such as in NMR) was given by Schulman and Vazirani~\cite{schulman1999molecular}. 
They showed that it is possible to reach polarization of order unity using only a number of qubits that scales as $1/{\epsilon_{b}^2}$ for initial polarization $\epsilon_b\ll1$. 
This scheme, later named ``reversible algorithmic cooling'', 
was improved by adding contact with a heat bath to cool the qubits that were
heated during the process~\cite{boykin2002algorithmic}. 
This improved method --called ``heat-bath algorithmic cooling'' (HBAC) -- allows to keep pumping entropy out of the system to the heat-bath, after each irreversible entropy-compression step in an iterative way.  
Based on this idea, many algorithms have been designed to purify a set of qubits by removing entropy of a subset of them at the expense of increasing the entropy of others~\cite{fernandez2004algorithmic,moussa:2005,schulman2005physical,elias2006optimal,schulman2007physical,elias2011semioptimal,kaye:2007,brassard2014prospects}. 
Beyond the theoretical interest, experiments have demonstrated
proof-of-principle of the reversible algorithm~\cite{chang2001nmr} and heat-bath
algorithm~\cite{fernandez2005paramagnetic,baugh2005experimental,Ryan:2008qf,elias2011heat,brassard2014prospects,park2014hyperfine},
and showed improvement in polarization for a few qubits.

The limits of HBAC have been studied using a specific algorithm, 
the Partner Pairing
Algorithm (PPA), which was introduced by Schulman, Mor and Weinstein~\cite{schulman2005physical},
and believed to be optimal among all possible HBAC~\cite{schulman2007physical}.
In the above two papers, they claimed that the PPA gives the optimal
physical cooling in terms of entropy extraction. However, the proof 
implicitly assumes that the qubits can only be independently refreshed with the bath (``qubit-reset''). 

In this paper, we use a more general reset operation, one that includes
correlated-qubits interaction with the bath. Interestingly, it  allows for better cooling
than the case where no correlations are present.
This is in agreement with recent work that has shown that quantum
correlations can play a role in work extraction and entropy flows in cooling protocols \cite{frey2014strong,brunner2014entanglement,perarnau2015extractable,rodriguez2015comments,liuzzo2016thermodynamics,PhysRevE.93.042135,goold2016role,dillenschneider2009energetics}. 

We present algorithms which lead to an increase purity
beyond the achievable cooling of PPA, 
as the coupling to the environment is not limited to independent qubit-relaxation.
In our model, we remove this restriction and include correlation between
qubits as they reset, also called cross relaxation.   It turns out
that our first algorithm is analogous to the Nuclear Overhauser Effect
(NOE)~\cite{Overhauser:1953fk}. NOE was discovered in 1953 and
describes the increase polarization of nuclear spins in presence of
decoupling and cross relaxation. We generalize our algorithm to the
case where we can control both the cross relaxations and
 independent qubit-relaxations and show an improvement on cooling.
We analytically calculate the maximum polarization achieved for the our method 
as a function of the number of qubits, 
and as a function of the heat-bath polarization.
We also find the polarization evolution as a function of the number of iterations of the algorithm and make a comparison with PPA.
We give explicit examples for two, three qubits and $n$ qubits, with analytical and numerical solutions for low and general bath polarization. 

\section{II. Background}
\begin{figure}[ht] 
\includegraphics[width=0.6\linewidth]{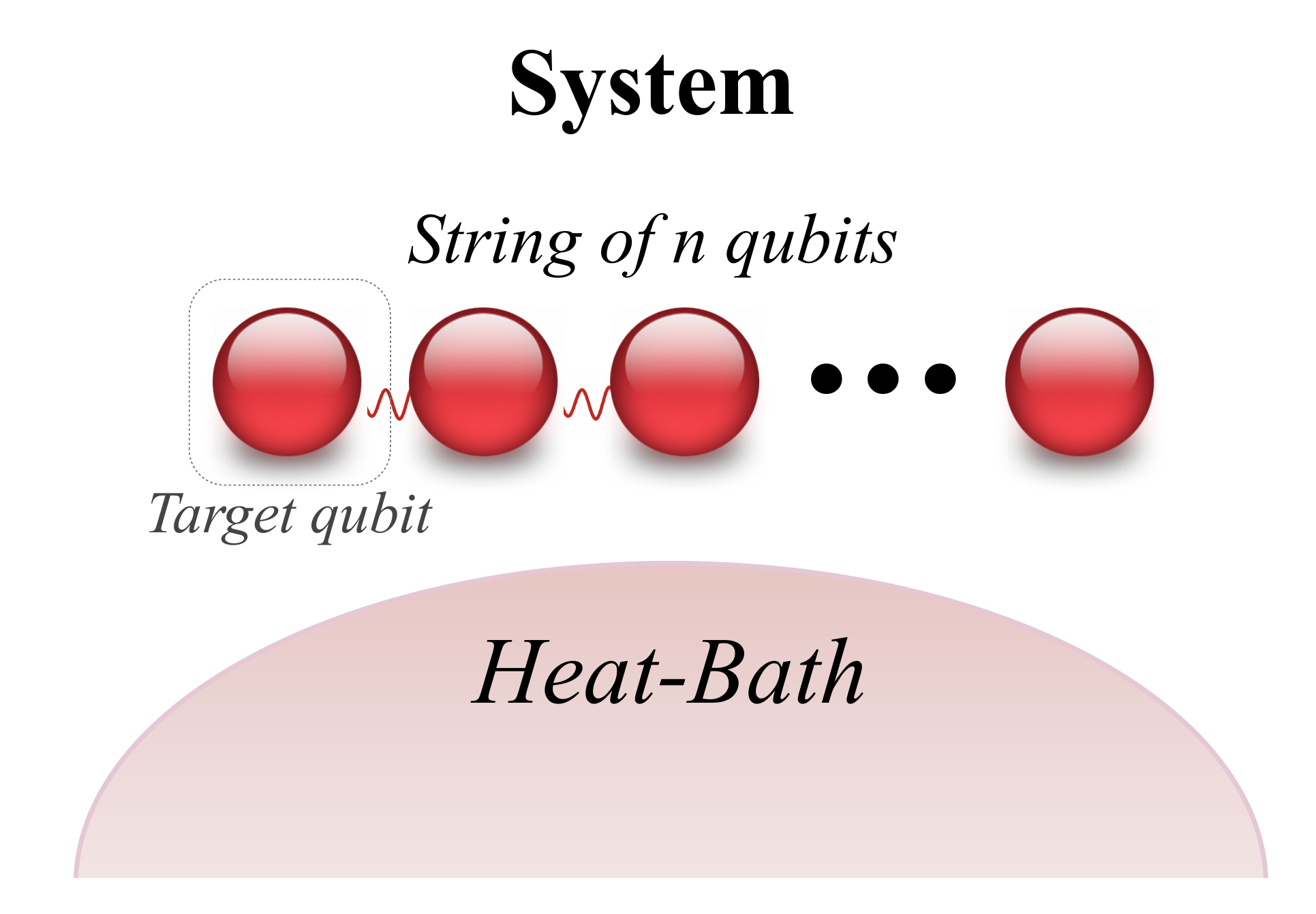}
\caption{The system consists of a string of $n$ qubits. The goal is to increase the purity of one of them (called the ``target qubit'') by a controlled redistribution of entropy among them, and pumping entropy out of the system into a heat-bath of polarization $\epsilon_b$.}
\label{fig:system.pdf}
\end{figure}

The system consists of a string of $n$ qubits: one qubit which is going to be cooled (also called the ``target qubit'') and $n-1$ qubits which aid in the entropy compression. We assume that the qubits can be brought into thermal contact with a heat-bath of polarization $\epsilon_b$ (Fig.\ref{fig:system.pdf}).

The PPA method purifies the target qubit by applying alternating rounds of entropy
compression and pumping entropy into a thermal bath of partially polarized qubits through qubit-reset operations (see a review in~\cite{park2015heat}).

The reversible entropy compression operation on a
string of $n$ qubits redistributes the entropy among the string. This is a unitary process that extracts entropy from the target qubit and concentrates that entropy in the reset qubits of the system.
In particular, PPA starts re-thermalizing all the qubits individually with the bath, before applying the rounds. That preliminary preparation leaves the system in a product state of all the qubits of the string with the same bath-polarization. For this case, all entropy compression steps will correspond to making a descending ordering (SORT) of the diagonal elements of the system's density matrix.
This operation aims to cool the first qubit, at the expense of warming the reset qubits. 

In the qubit-reset step, the PPA individually relaxes the qubits using the bath, which is equivalent to swapping the reset qubits with qubits from the heat-bath. In this paper, we introduce a more efficient way of extracting entropy from the system into the bath by coupling the quantum system with the bath.

The limits of the PPA method lead to an asymptotic polarization on the first qubit.
%
An exact steady state of the cooling limit of PPA
was recently found and presented 
in~\cite{PhysRevLett.116.170501,raeisi2015asymptotic}.
Using only one reset qubit and a total of $n$ qubits, PPA gives an asymptotic polarization~\cite{PhysRevLett.116.170501} of
\begin{equation}
\epsilon_{max}=\frac{(1+\epsilon_b)^{2^{n-2}}-(1-\epsilon_b)^{2^{n-2}}} {(1+\epsilon_b)^{2^{n-2}}+(1-\epsilon_b)^{2^{n-2}}}
\end{equation}
where $\epsilon_b$ is the polarization of the bath. 

This claim means that for a system of two qubits ($n=2$),
starting in the maximally mixed state, the PPA 
gives a steady state with the qubits at the bath temperature and no polarization gain (beyond that of the bath) is observed. 
However, in a recent paper~\cite{li2014maximally}, Li et al. studied the efficiency of polarization transfer in the presence of a bath
using {\it vector of coherence representation}.   
They showed a numerical solution and an experiment where the polarization enhancement is above the bath polarization for two qubits, 
thus bypassing the optimal cooling allowed by PPA. 
The surprising improvement turns out to be related 
to the Nuclear Overhauser Effect,
discovered in 1953~\cite{Overhauser:1953fk} 
(see also~\cite{brassard2014prospects} 
where NOE is mentioned in the context of HBAC, yet the bypass was not noted in
that work). 


%
Below, we show for the first time how to describe NOE using
quantum information processing terminology.

In the following sections, we show how to generalize the independent
reset of qubits with an environment to a ``state-reset'' operation in
quantum-information-processing terminology for $n$ qubits. This type of reset
provides HBAC-methods a way to pump entropy out of the system by
taking advantage of qubit-environment correlations.  Furthermore, we
not only describe NOE with a quantum algorithm, but also we
show an improved algorithm that goes beyond the polarization enhancement obtainable by NOE, and by the PPA class.


\section{III. Reset of states by correlated-qubits relaxation}

\subsection{A. State-reset $\Gamma_2$ (for $n=2$)}

\begin{figure}[ht] 
\includegraphics[width=0.5\linewidth]{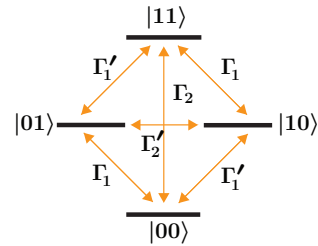}
\caption{Relaxation diagram for a two qubit system. The process
  illustrated as $\Gamma_2$ ($\ket{00}\leftrightarrow\ket{11}$) cannot be described as a qubit-reset with the bath as in the PPA, and results in the boost of polarization of one qubit when the other is saturated.}
\label{fig:relaxation}
\end{figure}

Consider a system of two qubits (Fig.\ref{fig:system.pdf}, $n=2$), and let $N_{00}$, $N_{01}$, $N_{10}$ and $N_{11}$ be the occupation numbers corresponding to the states $|00\rangle$, $|01\rangle$, $|10\rangle$ and $|11\rangle$, respectively.
Let $\Gamma_1$, $\Gamma'_1$,
$\Gamma_2$, and $\Gamma'_2$ be the transition probabilities per unit time between the four states, as depicted in Fig.~\ref{fig:relaxation}.
Note that the effect of $\Gamma_1$ is the usual single qubit relaxation, which corresponds to the ``reset-qubit".
However, when the rate associated with the double quantum transition, $\Gamma_2$, is larger than the other transition rates, the states $|00\rangle$ and $|11\rangle$ will relax towards equilibrium faster than the other states.
In the extreme case, when $\Gamma_2\neq0$ and the other transition probabilities equal to zero, there will be an equilibration between $N_{00}$ and $N_{11}$, while $N_{01}$ and $N_{10}$ will stay constant during the relaxation process. 
The Kraus operators which describe the evolution under $\Gamma_2$, are as follows:
\begin{equation}
\label{eq:Kraus_n2} 
\begin{split}
A^{\left(n=2\right)}_1 & =\sqrt{p_2} |00\rangle \langle 00 |, \\
A^{\left(n=2\right)}_2 & =\sqrt{p_2} |00\rangle \langle 11 |, \\
A^{\left(n=2\right)}_3 &=\sqrt{1-p_2} |11\rangle \langle 11 |, \\
A^{\left(n=2\right)}_4 & =\sqrt{1-p_2} |11\rangle \langle 00 |,\\
A^{\left(n=2\right)}_5 & =|01\rangle \langle 01 |, \\
A^{\left(n=2\right)}_6 & =|10\rangle \langle 10 |,
\end{split}
\end{equation}
where the probability $p_2$ is the sum of the probabilities of $|00\rangle$ and $|11\rangle$. This probability depends on the heat bath polarization $\epsilon_b$ as $p_2=\frac{e^{2\xi_b}}{2\cosh{2\xi_b}}$, where $\epsilon_b=\tanh{\xi_b}$. Thus, for low polarization, $p_2=\left(1+2\epsilon_b\right)/2$. 

We will use this case, because when $\Gamma_1$, $\Gamma'_1$ and $\Gamma'_2$ are larger than zero during the $\Gamma_2$ process, they can reduce the polarization enhancement.

The state of the system $\rho$, under the effect of $\Gamma_2$, will evolve to
\begin{equation}
\label{eq:Kraus_operatorsss} 
\Gamma_2\left(\rho\right)=\displaystyle \sum_{i=1}^{6}A_i^{\left(n=2\right)}\rho \left(A_i^{ \left(n=2\right)}\right)^{\dagger}.\end{equation}
This operation transforms the diagonal elements of the density matrix $\rho$, from  
$\operatorname{diag}\left(\rho\right)=\left(N_0, N_1,N_2, N_3\right)$ to
$\left(\left(N_0+N_3\right)p_2, N_1,N_2, \left(N_0+N_3\right)\left(1-p_2\right)\right)$, resulting in the state-reset between $|00\rangle$ and $|11\rangle$.

In general, we can reset not only the states  $|00\rangle$ and $|11\rangle$, but any pair of states, even in higher dimension (see section  ``Reset of states $\Gamma_n$" for further details about the required Kraus operators).

This form of reset, accompanied by saturation of the second qubit, can boost the polarization of the first qubit, as explained in the next section.

\subsection{B. Nuclear Overhauser Effect in QIP}

NOE emerges in a pair of qubits in the presence of cross-relaxation, when one of the qubits is rapidly rotated so that over relevant timescale its polarization averages to zero, resulting in the boost of polarization of the other qubit~\cite{Overhauser:1953fk}.

\begin{figure}[ht]
\includegraphics[width=0.45\textwidth]{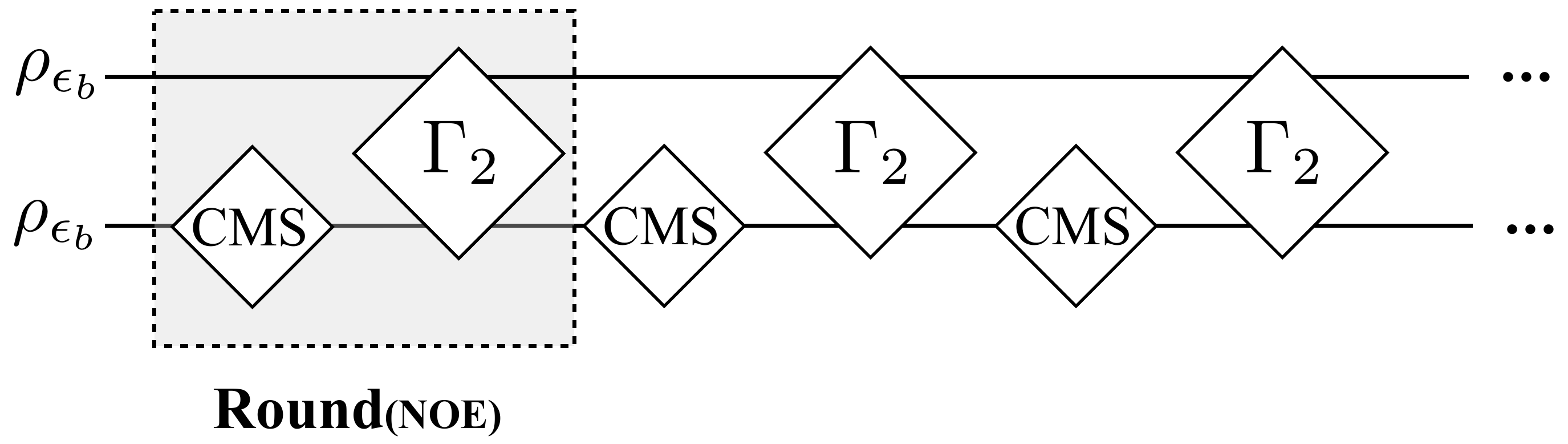}
\caption{
Quantum Circuit for NOE. $\Gamma_2$ is the state-reset operations obtained from cross relaxation, and $CMS$ the completely-mixed-state operation. The part inside the dotted box represents a round. The entire circuit is the repetition of that round.}
\label{fig:NOE}
\end{figure}

We can describe this process, in the quantum information processing terminology, as the iteration of two operations (see fig.~\ref{fig:NOE}), one to saturate (totally mix) the second qubit, using a gate that we call ``reset to completely-mixed-state, $CMS$"; and the other is $\Gamma_2$, for the relaxation of the system. We assume control over non-unitary processes, in the spirit of reservoir engineering (as in \cite{PhysRevLett.77.4728}).

Consider a system of two homonuclear spins. Let the system start in thermal equilibrium with polarization $\epsilon_b$, hence, the initial state is 
 $\rho^{\otimes 2}_0= \frac{1}{2} \begin{bmatrix}
 1+\epsilon_b && 0\\
0 &&  1-\epsilon_b
 \end{bmatrix} ^{\otimes 2}$. This state can be represented by the diagonal of its density matrix, $\operatorname{diag}\left(\rho^{\otimes 2}_0\right)$. For low polarization (see Appendix B for \emph{any} polarization), this diagonal is approximately  $\frac{1}{4}\left(1+2\epsilon, 1, 1, 1-2\epsilon_b\right)$. This state will evolve as follows, under two iterations of the two mentioned steps:
\begin{equation}
\label{eq:NOE_evo} 
\begin{split}
\operatorname{diag}\left(\rho^{\otimes 2}_0\right)=\frac{1}{4}\left(1+2\epsilon_b, 1, 1, 1-2\epsilon_b\right)\\
\xrightarrow{CMS(2)} \frac{1}{2}\left(1+\epsilon_b,1-\epsilon_b\right)\otimes\frac{1}{2}\left(1,1\right)\\
= \frac{1}{4}\left(1+\epsilon_b, 1+\epsilon_b, 1-\epsilon_b, 1-\epsilon_b\right)\\
\xrightarrow{\Gamma_2} \frac{1}{4}\left(1+2\epsilon_b,1+\epsilon_b,1-\epsilon_b,1-2\epsilon_b\right)\\
\xrightarrow{CMS(2)} \frac{1}{4}\left(1+\frac{3}{2}\epsilon_b,1-\frac{3}{2}\epsilon_b\right)\otimes \frac{1}{2}\left(1,1\right)\\
= \frac{1}{4}\left(1+\frac{3}{2}\epsilon_b, 1+\frac{3}{2}\epsilon_b, 1-\frac{3}{2}\epsilon_b, 1-\frac{3}{2}\epsilon_b\right)\\
\xrightarrow{\Gamma_2} \frac{1}{4}\left(1+2\epsilon_b,1+\frac{3}{2}\epsilon_b,1-\frac{3}{2}\epsilon_b,1-2\epsilon_b\right),\\
\end{split}
\end{equation}
 where $CMS(2)$ stands for the saturation process of the second qubit.
 
From eq.(\ref{eq:NOE_evo}), we can see that the polarization of the first qubit has an enhancement of $3/2$ after the first round, and of $7/4$ after the second round, and so on. This enhancement grows asymptotically to a fixed point, corresponding to the polarization $\epsilon^{\infty}_{NOE}$. I.e., in the limit of the algorithm, after applying an iteration,
\begin{equation}
\label{eq:NOE_fix} 
\begin{split}
\frac{1}{2}\left(1+\epsilon^{\infty}_{NOE},1-\epsilon^{\infty}_{NOE}\right)\otimes\frac{1}{2}\left(1,1\right)\\
\xrightarrow{\Gamma_2} \frac{1}{4}\left(1+2\epsilon_b,1+\epsilon^{\infty}_{NOE},1-\epsilon^{\infty}_{NOE},1-2\epsilon_b\right)\\
\xrightarrow{CMS} \frac{1}{2}\left(1+\frac{2\epsilon_b+\epsilon^{\infty}_{NOE}}{2},1-\frac{2\epsilon_b+\epsilon^{\infty}_{NOE}}{2}\right)\otimes \frac{1}{2}\left(1,1\right)
\end{split}
\end{equation}
the polarization of the first qubit should be the same. Thus,
$\epsilon^{\infty}_{NOE}=\left(2\epsilon_b+\epsilon^{\infty}_{NOE}\right)/2$,
leading to $\epsilon^{\infty}_{NOE}=2\epsilon_b$, the same enhancement obtained
from the Solomon equations as shown in Appendix A.  See precise calculation in Appendix B. Appendix C shows a generalization of NOE to $n$ qubits. Appendix D introduces a practical NOE-based
algorithm, leading to a different generalization of the (standard) NOE to $n$ qubits, a better cooling than the one described in Appendix C, and furthermore it provides an improvement over the PPA.

\subsection{C. State-Reset HBAC for the two-qubit case: SR$\boldsymbol{\Gamma_2}$-HBAC}

Now we present a different and better cooling algorithm which uses the
state-reset tool, beginning with the two-qubit state. This algorithm is related to what is known as transient NOE.
Let's start with two qubits at thermal equilibrium, with polarizations $\epsilon_b$, i.e. with state
\begin{equation}
\begin{split}
\rho^{n=2}_0=\rho_{\epsilon_b}^{\otimes 2}
&=\left[  \frac{1}{2}
\begin{pmatrix}
 1+\epsilon_b  & 0\\
 0 & 1-\epsilon_b\\
 \end{pmatrix}
\right]^{\otimes 2}\\
&=\left[  \frac{1}{2\cosh {\xi_b}}
\begin{pmatrix}
 e^{\xi_b}  & 0\\
 0 & e^{-\xi_b}\\
 \end{pmatrix}
\right]^{\otimes 2},
\end{split}
\label{edo_1}
\end{equation}
where $\epsilon_b=\operatorname{tanh}\left(\xi_b\right)$. This state can be expressed as a vector of its diagonal elements, $\operatorname{diag}\left(\rho^{n=2}_0\right)=\frac{1}{4\cosh^2{\xi_b}}\left(e^{2\xi_b},1,1,e^{-2\xi_b}\right)$.
\begin{figure}[h]
\includegraphics[width=0.49\textwidth]{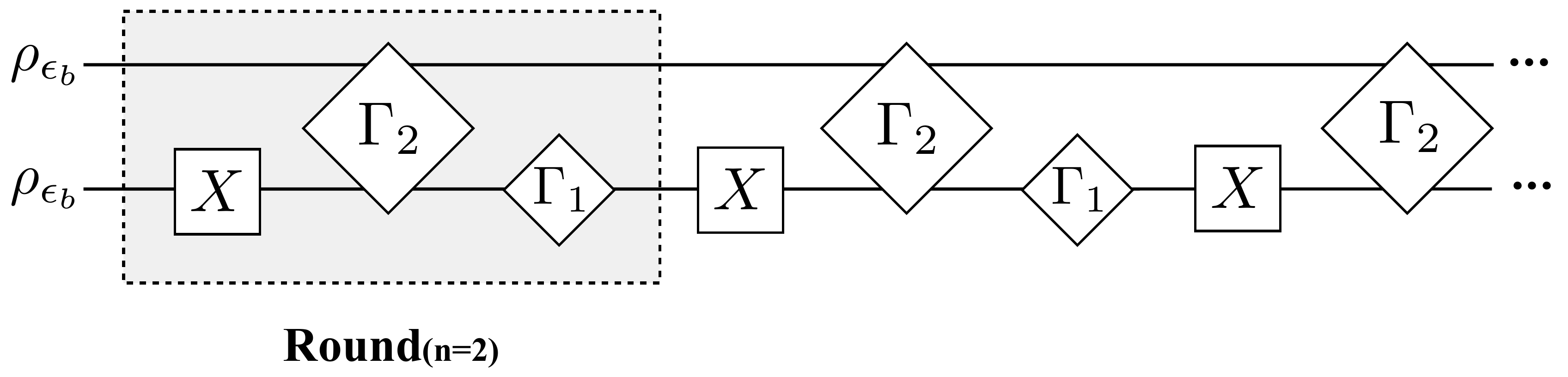}
\caption{
Circuit for the SR$\Gamma_2$-HBAC method. $\Gamma_1$ and $\Gamma_2$  are the state-reset operations on one and on two qubits, respectively. The part inside the dotted box represents a round, the entire circuit is just the repetition of that round.}
\label{fig:circuitn2new}
\end{figure}

The circuit of this algorithm is presented in Fig.~\ref{fig:circuitn2new}. We assume control over non-unitary processes, in the spirit of reservoir engineering (as in \cite{PhysRevLett.77.4728}) to create the effect of turning on $\Gamma_1$ and $\Gamma_2$ independently. Recall that $\Gamma_1$ returns a qubit to its corresponding initial equilibrium state, and $\Gamma_2$ transforms the diagonal of a state, $\rm{diag}\left(\rho\right)=\left(N_0, N_1, N_2, N_3\right)$, to 
$\left(\left(N_0+N_3\right)p_2, N_1, N_2,\left(N_0+N_3\right)\left(1-p_2\right)\right)$, where $p_2=\frac{e^{2\xi}}{2\cosh{2\xi}}$. 

For low polarization, the initial state will evolve from $\rm{diag}\left(\rho_0^{n=2}\right)\approx\frac{1}{4}\left(1+2\epsilon_b,1,1,1-2\epsilon_b\right)$ to $\rm{diag}\left(\rho_1\right)=\frac{1}{4}\left(1+2\epsilon_b,1+2\epsilon_b,1-2\epsilon_b,1-2\epsilon_b\right)$ after applying the first two gates of a round (see Fig.~\ref{fig:circuitn2new}). At this stage, the polarization of the first qubit is doubled and the second qubit  has zero polarization, analogous to the NOE~\cite{Overhauser:1953fk}. The third gate prepares the second qubit for the next round. By iterating the round, we can keep increasing its polarization: after applying $k$ rounds, it will be
\begin{equation}
 \epsilon^{\left(n=2\right)}_k=\left(3-\frac{1}{2^{k-1}}\right)\epsilon_b,
\label{pol_k_2q}
\end{equation}
for the low polarization case.
This leads to an asymptotic polarization of $3\epsilon_b$, an improvement on both the NOE and the PPA for the two-qubit case.

For a general polarization, the state-evolution during the first round of the algorithm will be as follows:
\begin{equation}
\label{eq:n2_ini_evo} 
\begin{split}
\operatorname{diag}\left(\rho^{n=2}_0\right)=\frac{1}{4\cosh^2{\xi_b}}\left(e^{2\xi_b},1,1,e^{-2\xi_b}\right)\\
\xrightarrow{X} \frac{1}{4\cosh^2{\xi_b}}\left(1,e^{2\xi_b},e^{-2\xi_b},1\right)\\
\xrightarrow{\Gamma_2}\frac{1}{4\cosh^2{\xi_b}}\left(2p_2,e^{2\xi_b},e^{-2\xi_b},2\left(1-p_2\right)\right)\\
= \frac{1}{4\cosh^2{\xi_b}}\left(\frac{e^{2\xi_b}}{\cosh{2\xi_b}},e^{2\xi_b},e^{-2\xi_b},\frac{e^{-2\xi_b}}{\cosh{2\xi_b}}\right)\\
\xrightarrow{\Gamma_1} \frac{1}{2\cosh{2\xi_b}}\left(e^{2\xi_b},e^{-2\xi_b}\right)\otimes\frac{1}{2}\left(p_1,1-p_1\right)\\
=\frac{1}{2\cosh{2\xi_b}}\left(e^{2\xi_b},e^{-2\xi_b}\right)\otimes\frac{1}{2\cosh{\xi_b}}\left(e^{\xi_b},e^{-\xi_b}\right).
\end{split}
\end{equation}
After the first round, the polarization of the first qubit will increase to $\tanh{2\xi_b}$ ($\approx 2\epsilon_b$ for low polarization).
Assume that after $k$ rounds of the algorithm, the polarization of the first qubit is $\epsilon^{\left(n=2\right)}_k=\tanh{\xi_k}$. Note that after the operation $\Gamma_1$, the system will be in a product state, 
with the second qubit with polarization $\epsilon_b=\tanh{\xi_b}$. 
Then, we apply one more round to obtain $\epsilon^{\left(n=2\right)}_{k+1}$, as follows:
\begin{equation}
\label{eq:new_algo_n2_evo0_low} 
\begin{split}
&\frac{1}{2\cosh{\xi_k}}\left(e^{\xi_k},e^{-\xi_k}\right)\otimes\frac{1}{2\cosh{\xi_b}}\left(e^{\xi_b},e^{-\xi_b}\right)=\\
&\frac{1}{4}\left(1+\tanh{\xi_k},1-\tanh{\xi_k}\right)\otimes\left(1+\tanh{\xi_b},1-\tanh{\xi_b}\right)\\
&\xrightarrow{X}\frac{1}{4}\left(1+\tanh{\xi_k},1-\tanh{\xi_k}\right)\otimes\\
&\quad  \quad \left(1-\tanh{\xi_b},1+\tanh{\xi_b}\right)\\
&\xrightarrow{\Gamma_2}\frac{1}{4}(\left(2-2\tanh{\xi_k}\tanh{\xi_b}\right)p_2,\\
&\quad  \quad\left(1+\tanh{\xi_k}\right)\left(1+\tanh{\xi_b}\right),\\
&\quad  \quad\left(1-\tanh{\xi_k}\right)\left(1-\tanh{\xi_b}\right),\\
&\quad  \quad\left(2-2\tanh{\xi_k}\tanh{\xi_b}\right)\left(1-p_2\right))\\
&=\quad\frac{1}{4}(\left(1-\tanh{\xi_k}\tanh{\xi_b}\right)\left[1+\tanh (2 \xi_b)\right],\\
&\quad  \quad\left(1+\tanh{\xi_k}\right)\left(1+\tanh{\xi_b}\right),\\
&\quad  \quad\left(1-\tanh{\xi_k}\right)\left(1-\tanh{\xi_b}\right),\\
&\quad  \quad\left(1-\tanh{\xi_k}\tanh{\xi_b}\right)\left[1-\tanh (2 \xi_b)\right])\\
&\xrightarrow{\Gamma_1}\frac{1}{2}(1+\frac{1}{2} \text{sech}(2 \xi_b)\left[\sinh (3 \xi_b) \text{sech}\xi_b+\tanh\xi_k\right],\\
&\quad  \quad 1-\frac{1}{2} \text{sech}(2 \xi_b)\left[\sinh (3 \xi_b) \text{sech}\xi_b+\tanh\xi_k\right])\otimes\\
&\quad\quad \frac{1}{2}\left(1+\tanh{\xi_b},1-\tanh{\xi_b}\right).
\end{split}
\end{equation}

\begin{figure}[h]
\centering
\includegraphics[width=0.48\textwidth]{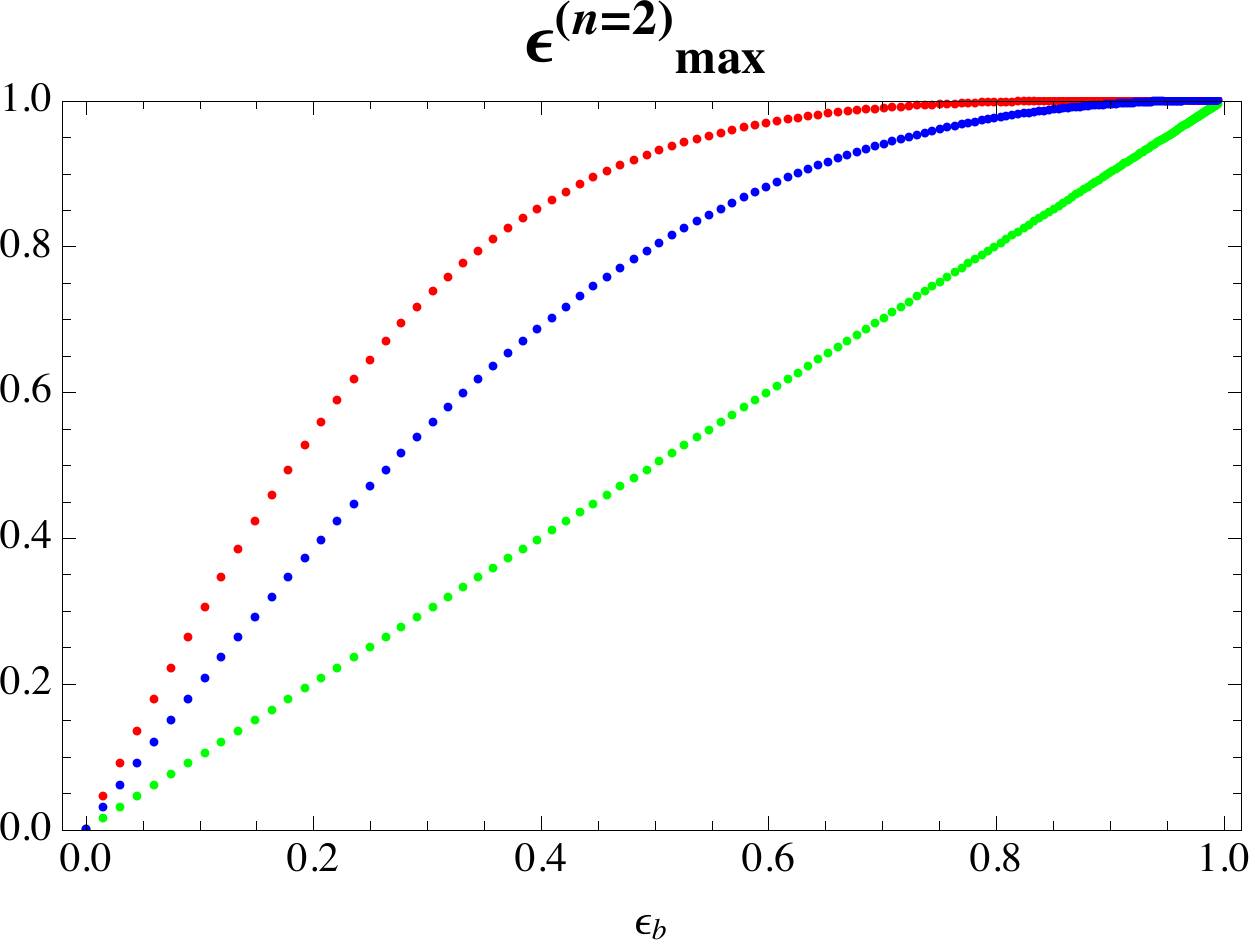}
\caption{
Maximum polarization achievable for various cooling  algorithms for the
two qubit case, assuming thermal equilibrium polarization $\epsilon_b$.  In red, results obtained by simulation of our algorithm presented in the circuit
of Fig.~\ref{fig:circuitn2new}, in blue, the simulation when the effect of $\Gamma_1$ is not included in our algorithm (analogous to the NOE) and 
in green, using the PPA algorithm.}
\label{fig:finalpol2}
\end{figure}

From here, the polarization increases from $\epsilon_{k}$ to $\epsilon_{k+1}=\frac{1}{2} \text{sech}(2 \xi_b)\left[\sinh (3 \xi_b) \text{sech}\xi_b+\tanh\xi_k\right]$. With initial polarization $\epsilon_0=\epsilon_b$ $(=\tanh\xi_b)$, we find that $\epsilon_{k}\leq\epsilon_{k+1}$, for all $k$.
In the cooling limit, $\epsilon_{\infty}=\epsilon_{\infty+1}$, i.e. 
\begin{equation}
\tanh{\xi_\infty=\frac{1}{2} \text{sech}(2 \xi_b)\left[\sinh (3 \xi_b) \text{sech}\xi_b+\tanh\xi_\infty\right]}.
\end{equation} 
From here, the maximum polarization achievable, in the two-qubit case starting with polarization $\epsilon_b$, will be
\begin{equation}
 \epsilon^{\left(n=2\right)}_\infty=\tanh{3\xi_b},
\label{maxpoln2}
\end{equation}
leading to an improvement on both the NOE and the PPA. 
For low polarization, $\epsilon_{k+1}\approx\frac{3\epsilon_b+\epsilon_k}{2}$. As $\epsilon_0=\epsilon_b$, the evolution of the first qubit polarization after applying $k$ rounds, will result in eq.~\ref{pol_k_2q}, leading to an asymptotic polarization of $3\epsilon_b$. See Fig.~\ref{fig:finalpol2} for numerical simulation results comparing PPA, NOE and this algorithm for the two-qubit case.

The control over non-unitary processes to turn on/off $\Gamma_1$ and $\Gamma_2$ independently is needed to have them individually, because the presence of $\Gamma_1$ at the same time as $\Gamma_2$ will decrease the polarization enhancement.


\section{IV. Generalized reset of states by correlated-qubits relaxation}

\subsection{A. State-reset $\Gamma_n$}

It is possible to generalize the state-reset idea to increase the polarization of a qubit in a larger $n-$qubit system by considering the transition probability between the eigenstates $|0\rangle^{\otimes n}$ and  $|1\rangle^{\otimes n}$, $\Gamma_n$. Similarly, in the limit case, when $\Gamma_n\neq0$, and the other transition probabilities equal to zero, the cross relaxation effectively provides a state-reset between  $|0\rangle^{\otimes n}$ and $|1\rangle^{\otimes n}$, without changing the other states. The Kraus operators that reset these states are as follows:
\begin{equation}
\label{eq:Kraus_n} 
\begin{split}
A^{\left(n\right)}_1 & =\sqrt{p_n}  \displaystyle \sbigotimes_{i=1}^n |0\rangle_i \langle 0 |_i, \\
A^{\left(n\right)}_2 & =\sqrt{p_n} \displaystyle \sbigotimes_{i=1}^n |0\rangle_i \langle 1 |_i,, \\
A^{\left(n\right)}_3 &=\sqrt{1-p_n} \displaystyle \sbigotimes_{i=1}^n |1\rangle_i \langle 1 |_i,, \\
A^{\left(n\right)}_4 & =\sqrt{1-p_n} \displaystyle \sbigotimes_{i=1}^n |1\rangle_i \langle 1 |_i,,\\
A^{\left(n\right)}_{4+j} & =\left(|j_{_{bin}}\rangle \langle j_{_{bin}} |\right),
\end{split}
\end{equation}
for all $j$ integer numbers between 0 and $2^n-1$, where the subscript $i$ correspond to the $i$-th qubit, and $j_{_{bin}}$ is the binary representation of $j$ written as a string of $n$ digits. (For instance, for $n=3$,  $A^{\left(n=3\right)}_{5}=\left(|001\rangle \langle 001 |\right)$, $A^{\left(n=3\right)}_{6} =\left(|010\rangle \langle 010 |\right)$, ..., and $A^{\left(n=3\right)}_{10}=\left(|110\rangle \langle 110 |\right)$).

In general, the operation to reset any pair of states, $|s_1\rangle$ and $|s_2\rangle$), can be given by a Kraus decomposition with the following operators: 
\begin{equation}
\label{eq:Kraus_g_n} 
\begin{split}
A^{\left(n\right)}_{1_{\left(|s_1\rangle \longleftrightarrow |s_2\rangle \right)}} & =\sqrt{p_n} |s_1\rangle \langle s_1|, \\
A^{\left(n\right)}_{2_{\left(|s_1\rangle \longleftrightarrow |s_2\rangle \right)}}  & =\sqrt{p_n} |s_1\rangle \langle s_2 |, \\
A^{\left(n\right)}_{3_{\left(|s_1\rangle \longleftrightarrow |s_2\rangle \right)}} &=\sqrt{1-p_n} |s_2\rangle \langle s_2 |, \\
A^{\left(n\right)}_{4_{\left(|s_1\rangle \longleftrightarrow |s_2\rangle \right)}} & =\sqrt{1-p_n} |s_2\rangle \langle s_1 |,\\
A^{\left(n\right)}_{{4+r}_{\left(|s_1\rangle \longleftrightarrow |s_2\rangle \right)}} & =\left(|r_{bin}\rangle \langle r_{bin} |\right),
\end{split}
\end{equation}
for all the $r_{bin}$ integer binary numbers of a string of $n$ digits such that $r_{_{bin}}\in\{0,1\}^{\otimes n}\backslash \{s_1,s_2\}$, and $r\in \mathbbm{N}$ is used for indexing the Kraus operators. When the state-reset is chosen to be between  $|0\rangle^{\otimes n}$ and $|1\rangle^{\otimes n}$, the probability $p_n$ is related to the heat bath polarization $\epsilon_b$ as $p_n=\frac{e^{n\xi_b}}{2\cosh{n\xi_b}}$, where $\epsilon_b=\tanh{\xi_b}$. 
Thus, for low polarization, $p_n=\left(1+n\epsilon_b\right)/2$.
This operation transforms the state $\rho$ to 
$\Gamma_n\left(\rho\right)= \sum_{i}{A^{\left(n\right)}_i}_{\left(|s_1\rangle \longleftrightarrow |s_2\rangle  \right)}\rho \left({A^{\left(n\right)}_i}_{\left(|s_1\rangle \longleftrightarrow |s_2\rangle \right)}\right)^{\dagger}
$.
%


Similarly, this kind of reset, combined with rotations to totally mix all the qubits with the exception of the qubit that is going to be cooled, can boost the polarization of some qubits. We found a generalization of the NOE on $n$ qubits (see appendix D), that gives a final polarization of
$n\epsilon_b$,  
 in the approximation of low polarization.
Used by itself, this generalized NOE does not always give a better polarization than the PPA for large $n$, but we use this state-reset operation as a tool to create new HBAC algorithms which give better results than just re-thermalizing the hot qubits and applying a unitary operator (as in the PPA). In the next section, we present our improved algorithm on $n$ qubits (starting with the three qubit case and then extending to $n$ qubits). We show the results for the maximum polarization achievable by our algorithm, and some comparisons with the PPA.


\subsection{B. State-Reset HBAC for the three-qubit case: SR$\boldsymbol{\Gamma_3}$-HBAC}

\begin{figure}[h]
\centering
\includegraphics[width=0.48\textwidth]{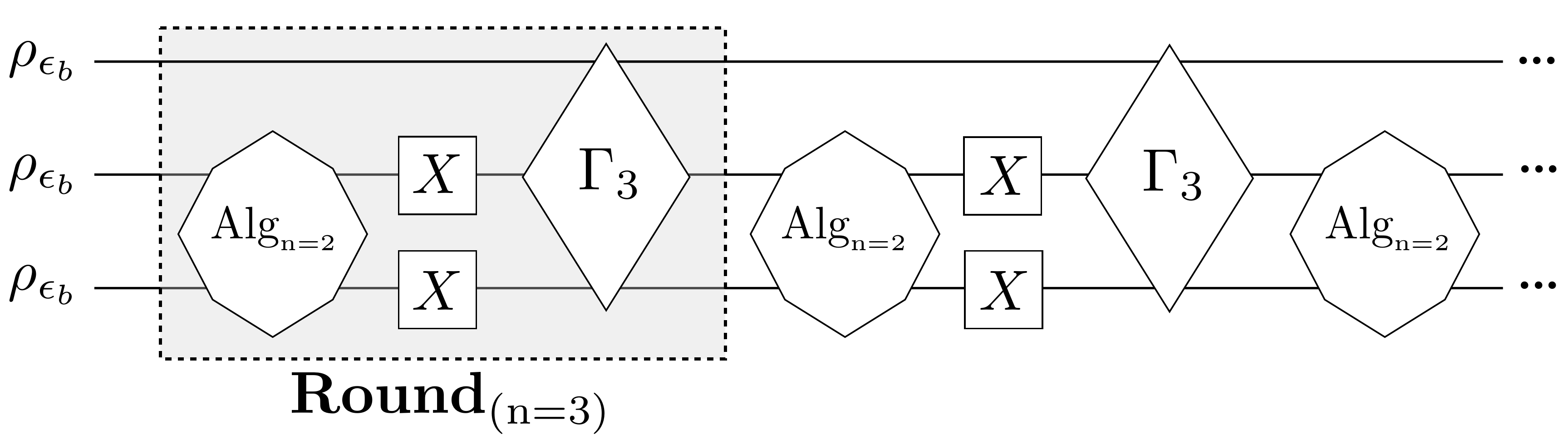}
\caption{
Circuit for the SR$\Gamma_3$-HBAC method. $\Gamma_3$ is the reset-state operation on three qubits that resets the states $|000\rangle$ and $|111\rangle$ to their corresponding equilibrium values. $\operatorname{Alg_{\left(n=2\right)}}$ is the preparation of the second qubit, using the SR$\Gamma_2$-HBAC on the second and third qubits. The dotted box encloses a segment of the circuit which is repeated: a $\rm{Round}_{\left(n=3\right)}$.}
\label{fig:circuitn3}
\end{figure}

To extend to three qubits we include $\Gamma_3$, in addition to $\Gamma_2$ which can be applied on any combination of qubit-pairs of the system, and $\Gamma_1$ on any of the qubits.
The algorithm consists of the iteration of three-step rounds (see Fig.~\ref{fig:circuitn3}). First, the polarization of the second qubit is increased by applying the SR$\Gamma_2$-HBAC on the second and third qubits, to achieve a polarization close enough to its maximum value (see eq.~(\ref{maxpoln2})), $\tanh{3\xi}$ ($3\epsilon_b$ for low polarization). 
Then, the second and third qubits are flipped. Finally, a state-reset $\Gamma_3$ on all three qubits is applied to pump additional entropy out of the system. 

Let's start with three qubits at thermal equilibrium with polarizations $\epsilon_b$, i.e. in the state
\begin{equation}
\begin{split}
\rho^{n=3}_0 &=\rho_{\epsilon_b}^{\otimes 3}\\
&=\left[  \frac{1}{2}
\begin{pmatrix}
 1+\epsilon_b  & 0\\
 0 & 1-\epsilon_b\\
 \end{pmatrix}
\right]^{\otimes 3}\\
&=\left[  \frac{1}{2\cosh {\xi_b}}
\begin{pmatrix}
 e^{\xi_b}  & 0\\
 0 & e^{-\xi_b}\\
 \end{pmatrix}
\right]^{\otimes 3},
\end{split}
\label{edo_1_2}
\end{equation}
where $\epsilon_b=\rm{tanh}\left(\xi_b\right)$. Then, the vector of the diagonal elements is $\operatorname{diag}\left(\rho^{n=3}_0\right)= \left[\frac{1}{2}\left(1+\epsilon_b,1-\epsilon_b\right)\right]^{\otimes 3}$.

For low polarization, $\Gamma_3$ will transform a diagonal state $\rm{diag} \left(\rho^{n=3}\right)=\left(N_0, N_1,N_2, N_3, N_4, N_5, N_6, N_7\right)$ 
into 
$\left(\frac{1+3\epsilon_b}{2}\left(N_0+N_7\right),N_1,N_2,N_3,N_4,N_5,N_6,\frac{1-3\epsilon_b}{2}\left(N_0+N_7\right)\right)$.
$\rm{Alg_{\left(n=2\right)}}$ will prepare the second qubit by applying the SR$\Gamma_2$-HBAC to enhance its polarization close to $3\epsilon_b$. Then, under this algorithm, the polarization of the first qubit, after applying $k'$ rounds $\rm{Round_{\left(n=3\right)}}$ (see Fig. \ref{fig:circuitn3}), will be 
\begin{equation}
 \epsilon^{n=3}_{k'}=\left[7-6\left(\frac{3}{4}\right)^{k'}\right]\epsilon_b,
\label{pol_k_3q}
\end{equation}
leading to an  asymptotic polarization $\epsilon^{\left(n=3\right)}_{max}=7\epsilon_b$.

In general for any $\epsilon_b$, to calculate the polarization's evolution as a function of the number of rounds, let $\epsilon^{\left(n=3\right)}_k$ be the polarization of the first qubit after $k$ rounds. Then, let's apply one more round on the system to obtain $\epsilon^{\left(n=3\right)}_{k+1}$.
After applying $\rm{Alg}_{n=2}$ in round $k+1$, 
the state will be
$\frac{1}{2}\left(1+\epsilon_k,1-\epsilon_k\right)\otimes\frac{1}{2}\left(1+\tanh{3\xi_b},1-\tanh{3\xi_b}\right)\otimes\frac{1}{2}\left(1+\epsilon_b,1-\epsilon_b\right)$. 
Then, by flipping the second and third qubit, the state will evolve to $\frac{1}{2}\left(1+\epsilon_k,1-\epsilon_k\right)\otimes\frac{1}{2}\left(1-\tanh{3\xi_b},1+\tanh{3\xi_b}\right)\otimes
\frac{1}{2}\left(1-\epsilon_b,1+\epsilon_b\right)$.
At this point, the first and last elements of the diagonal density matrix are $\alpha_1:=1/8(1+\epsilon_k)(1-\tanh{3\xi_b})(1-\epsilon_b)$ and $\alpha_2:=1/8(1-\epsilon_k)(1+\tanh{3\xi_b})(1+\epsilon_b)$, respectively. The sum of these two elements is $A:=\alpha_1+\alpha_2=1/4(1-\epsilon_k\tanh{3\xi_b}+\tanh{3\xi_b}\epsilon_b-\epsilon_k\epsilon_b)$, thus the state-reset $\Gamma_3$ will change these elements to $A p_3$ and to $A (1-p_3)$, respectively.
Thus, the new polarization of the first qubit will be $\epsilon_{k+1}=\epsilon_k+2(Ap_3-\alpha_1)$. 
Substituting $\alpha_1$, $A$, $p_3=(1+\tanh{3\xi_b})/2$, and $\epsilon_b=\tanh{\xi_b}$ in $\epsilon_{k+1}$, we obtain
\begin{equation}
 \epsilon^{\left(n=3\right)}_{k+1}=\frac{\epsilon^{\left(n=3\right)}_k\left(2\cosh{\xi_b}+\cosh{5\xi_b}\right)+\sinh{7\xi_b}}{2\cosh{\xi_b}+\cosh{5\xi_b}+\cosh{7\xi_b}}.
\label{pol_k+1_3q}
\end{equation}
From here, starting with polarization $\epsilon_{0}=\epsilon_b$ $(=\tanh{\xi_b})$, each round gives an improvement, $\epsilon_{k}\leq\epsilon_{k+1}$, for all $k$.

In the cooling limit it is not possible to keep increasing this purity, i.e. $\epsilon^{(n=3)}_{\infty}=\epsilon^{n=3}_{\infty+1}$, then, from eq.(\ref{pol_k+1_3q}), the maximum polarization achievable with our algorithm for the three qubit case is
\begin{equation}
 \epsilon^{\left(n=3\right)}_\infty=\tanh{7\xi_b},
\label{maxpoln3}
\end{equation}
leading to an improvement on both the NOE and the PPA. 
%


\subsection{C. State-Reset HBAC for the $n$-qubit case: SR$\boldsymbol{\Gamma_n}$-HBAC}

Assume that we have the ability to apply state-reset operations $\Gamma_m$ in a controlled way on any subsystem of $m$ qubits.
Similarly to the algorithm for the three qubit case SR$\Gamma_3$-HBAC, which makes use of the preparation SR$\Gamma_2$-HBAC to enhance the polarization in the two qubits case, here our algorithm for $n$ qubits, SR$\Gamma_n$-HBAC, uses the preparation of the $\left(n-1\right)$-qubit case, SR$\Gamma_{n-1}$-HBAC. 
Again, the algorithm consists of the iteration of three-step rounds. First, the polarization of the second qubit, is increased by using the preparation of the $\left(n-1\right)$-algorithm, $\rm{Alg_{\left(n-1\right)}}$. 
Second, all the qubits with the exception of the first qubit are inverted.
Finally, a state-reset operation $\Gamma_n$ is applied to pump entropy out of the first qubit, see Fig.~\ref{fig:circuitn3}.

\begin{figure}[h]
\includegraphics[width=0.48\textwidth]{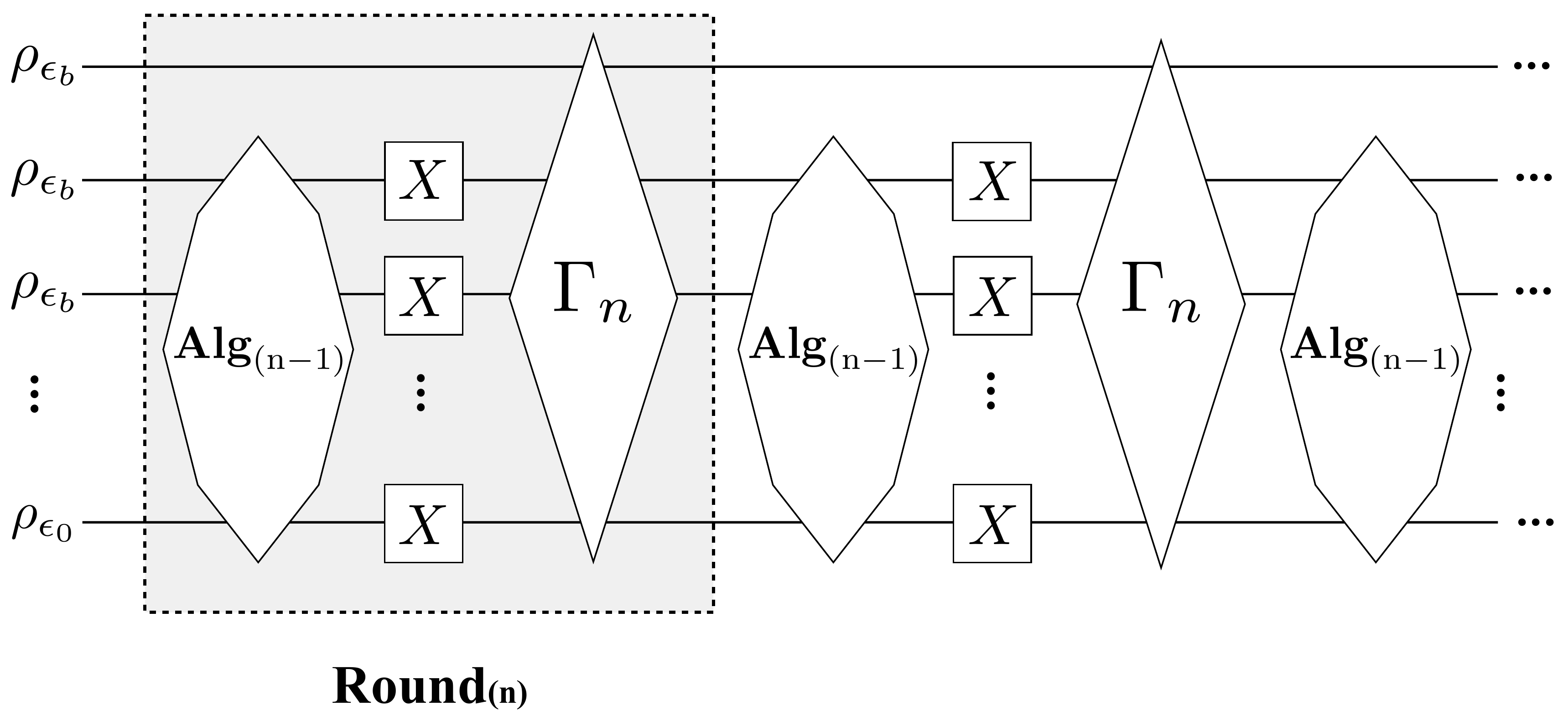}
\caption{
Circuit for the SR$\Gamma_n$-HBAC method. $\Gamma_n$ is the reset-state operation on $n$ qubits that resets the states $|0\rangle^{\otimes n}$ and $|1\rangle^{\otimes n}$. $\rm{Alg}_{n-1}$ is the polarization preparation of the second qubit by using SR$\Gamma_{n-1}$-HBAC. The part of the circuit inside the dotted box represents a round for $n$ qubits, $\rm{Round}_{(n)}$. The entire circuit is the repetition of this round.}
\label{fig:circuitnn}
\end{figure}

The analytical result for the maximum polarization of the first qubit, for any heat-bath polarization, will be
\begin{equation}
 \epsilon^{\left(n\right)}_\infty=\tanh{\left[(2^{n}-1)\xi_b\right]},
\label{max_pol_nq}
\end{equation}
where $\xi_b=\operatorname{arctanh}{\left(\epsilon_b\right)}$.

\begin{figure}[h] 
\center{\includegraphics[width=1\linewidth]{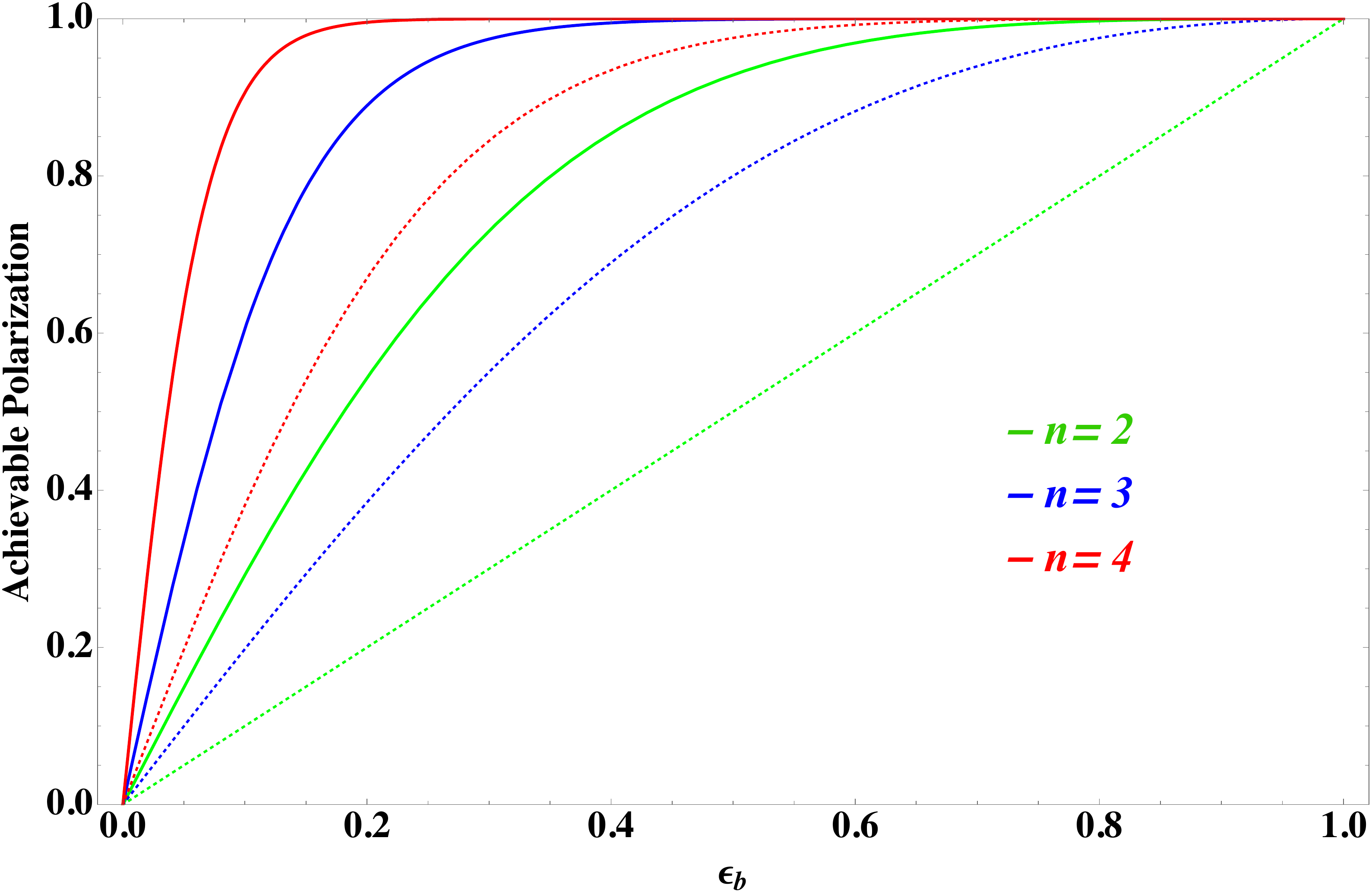}}
\caption{Maximum achievable polarizations for the SR$\Gamma_n$-HBAC method vs. the PPA, in solid lines and dotted lines, respectively, as a function of $\epsilon_b$, for different $n$.}
\label{fig:maxpol}
\end{figure}

This maximum polarization is in general higher than the achievable polarization obtained by the PPA method~\cite{PhysRevLett.116.170501}. 
Fig.~\ref{fig:maxpol} shows particular examples of the maximum polarizations by our method in comparison with the PPA, as a function of $\epsilon_b$. We found this maximum achievable polarization proved by induction, as described below.

The basis case of induction, for $n=2$ and $3$, the maximum polarization for the first qubit is $\epsilon^{(n=2)}_\infty=\tanh{3\xi_b}$ and $\epsilon^{(n=3)}_\infty=\tanh{7\xi_b}$, respectively.
In the induction step, we assume that $\epsilon^{(k)}_\infty=\tanh{\left(2^{k}-1\right)\xi_b},\, \forall k \leq \tilde{n}$, and prove this equation for $k=\tilde{n}+1$. 
Let's consider a system of $\tilde{n}+1$ qubits, we are going to calculate the maximum polarization $\epsilon^{(\tilde{n}+1)}_\infty$. 
After applying SR$\Gamma_n$-HBAC all the qubits will be in a product state, with the last $\tilde{n}$ qubits having the corresponding maximum polarization (i.e., the last qubit with polarization $\epsilon^{(n=1)}_\infty$, the second last one with polarization $\epsilon^{(n=2)}_\infty$, and so on). Let's name $\epsilon_{fix}^{\left(n+1\right)}$ to the polarization of the first qubit in the cooling limit. 
After the second step of the round, flipping the last $\tilde{n}$ qubits, the first element of the diagonal density matrix will be
$\displaystyle \beta_1:=\frac{1}{2^{n+1}}(1+\epsilon_{fix}^{\left(n+1\right)})\prod^n_{i=1}\left[1-\tanh{\left[(2^i-1)\xi_b\right]}\right]$.
Similarly, the last element of the density matrix will be
$\displaystyle \beta_{2^{n+1}}:=\frac{1}{2^{n+1}}(1-\epsilon_{fix}^{\left(n+1\right)})\prod^n_{i=1}\left[1+\tanh{\left[(2^i-1)\xi_b\right]}\right].$

Let's define the sum of these two elements, $\beta_1$ and $\beta_{2^{n+1}}$, as $B$. Then, the state-reset $\Gamma_{n+1}$ will change these two elements to $B p_{n+1}$ and to $B (1-p_{n+1})$, respectively.

This state will achieve the fixed point when the first and the last elements are already equal to $B p_{n+1}$ and to $B(1-p_3)$, respectively. Namely,
$\beta_1=Bp_{n+1}=\left(\beta_1+\beta_{2^{n+1}}\right)\left(1+\tanh{n\xi_b}\right)/2$. Substituting $\beta_1$ and $\beta_{2^{n+1}}$ in this expression, and solving for $\epsilon_{fix}^{\left(n+1\right)}$, we get $\epsilon_{fix}^{\left(n+1\right)}=\tanh{\left[(2^{n+1}-1)\xi_b\right]}$, which proves the claim: The polarization limit, achievable with our algorithm, for the $n$-qubit case is $\epsilon^{\left(n\right)}_\infty=\tanh{\left[(2^{n}-1)\xi_b\right]}$,
leading to an improvement on both the NOE and the PPA. For low polarization, this polarization limit reduces to $(2^{n}-1)\epsilon_b$

\section{V. Conclusion}

In conclusion, we have presented a new HBAC technique that can have a
better polarization enhancement than what had been achieved for the
PPA, for any number of qubits. Our technique utilizes the coupling to
the environment in a way that is not limited to qubit resets, but could also include correlations between the qubits which are reset (we named this ``state-reset'').
The assumption that entropy can be extracted from the system only via qubit reset (instead of state-reset) was a symmetry
which has been implicitly imposed for qubits, but is not
generally true.  There are other examples of similar imposed symmetry, such
as the distinction on subspace and subsystems~\cite{knill2000theory}
where the symmetry limits quantum information processing.
We have shown a series of algorithms and calculated their resulting
polarization for this new method as a function of the number of
qubits, $n$, and as a function of the polarization of the bath,
$\epsilon_b$. 
We have also presented the polarization evolution as a function of the
number of iterations of our algorithms and compared between these results and the corresponding ones of the PPA.
%
Our results show implicitly that a universal set of
unitary gates along with $\Gamma_1$ are not universal for open quantum systems.
We conjecture that including $\Gamma_2$ achieves universality. 
Future research should also include a proof of optimality of our
algorithms either using only the transitions $\ket{0}^{\otimes n} \longleftrightarrow \ket{1}^{\otimes n}$ (``state reset''), a different $m$-qubit transition (where $m\le n$), or a combination of such transitions.
The thermodynamics view of these algorithms
will explore in a subsequent publication.

\textit{Acknowledgments.} --- The authors would like to thank Xian Ma, Aharon Brodutch, Osama
Moussa, Daniel Park, David Cory, Om Patange, David Layden, and Miriam Gauntlett for
insightful discussions. N.~A.~R.-B. is supported by CONACYT-COZCYT and
SEP. R.~L. is supported by Industry Canada and the government of Ontario,
CIFAR. R.~L., T.~M. and Y.~W. thank the Schwartz/Reisman Foundation.


\section{Appendix A: The Nuclear Overhauser Effect}


Consider a system of two qubits. NOE appears in the presence of cross-relaxation and it  
results in a boost of polarization of one of the qubits when the second one is saturated, i.e. rotated
rapidly so that over relevant 
timescale its polarization averages to zero.  
This can be seen from the Solomon equation~\cite{Solomon:1955uq},
\begin{equation}
\frac{d \langle Z^1\rangle}{dt}= -R_1 (\langle Z^1\rangle - \langle Z^1\rangle_0)
-R_{12} (\langle Z^2\rangle - \langle Z^2\rangle_0),
\label{szevol}
\end{equation}
where $\langle Z^i\rangle$ is the polarization of the $i^{th}$ qubit,
$\langle Z^i\rangle_0$ the corresponding polarization in their equilibrium values,
$R_1=\Gamma'_2+2\Gamma_1+\Gamma_2$ and $R_{12}=\Gamma_2-\Gamma'_2$ are
combinations of relaxation rates $\Gamma_i$ for various transition as depicted in Fig.~\ref{fig:relaxation}. 
When $\langle Z^2\rangle=0$, the steady state solution is
\begin{equation}
\langle Z^1\rangle = \langle Z^1\rangle_0 + \frac{R_{12}}{R_1} \langle Z^2\rangle_0.
\end{equation}

This gives an enhancement beyond PPA for the two-qubit case, as long as $R_{12}/{R_1}$ is positive. In particular, when 
$\Gamma'_2=\Gamma_1=0$ we obtain an enhancement of $2$, for two spins with identical thermal-equilibrium polarizations.
The effect relies on cross-relaxation, and cannot be understood as a simple “swap” of the polarization of the reset qubits with the polarization of the bath.
In this paper, we show that one way to understand this process from an algorithmic point of view is to realize that the cross-relaxation effectively provides a state relaxation/equilibration between $|00\rangle $ and $|11\rangle$ (obtaining a ``state-reset''), and leaving other states intact. Furthermore, this state-reset accompanied by a rotation of the second qubit can boost the polarization of the first qubit beyond what would be obtained by a ``qubit-reset'' in the PPA.

\section{Appendix B: Precise calculation for NOE on two qubits}

We generalize our calculation to any bath-polarization, $0 < \epsilon_b < 1$, where $\epsilon_b = \tanh(\xi)$, following the circuit of Fig.\ref{fig:NOE}. The operation $\Gamma_2$, applied to an initial diagonal density-matrix of a 2-qubit-system, produces
\begin{eqnarray}
&&\rm{diag}\left(\rho\right)=\left(N_0,N_1,N_2,N_3\right)\nonumber\\
&&\xrightarrow{\Gamma_2}[(N_0+N_3)p_2,N_1,N_2,(N_0+N_3)(1-p_2)],
\label{eq:Nayeli}
\end{eqnarray}
where $p_2=\frac{e^{2\xi}}{2\cosh{2\xi}}$ is the population of the state $\left|00\right\rangle$ at thermal equilibrium with the heat-bath, normalized by the sum of thermal populations of both states $\left|00\right\rangle$ and $\left|11\right\rangle$ (i.e. $p_2=e^{2\xi}/N$, where $N=e^{2\xi}+e^{2\xi}=2\cosh2\xi$); and so $1-p_2$ is the complementary population of  $\left|11\right\rangle$ at thermal equilibrium.

In the cooling limit, the polarization of the target qubit $\epsilon^{\infty}_{NOE}$ is a fix point of the algorithm, thus after applying an iteration it will remain the same, i.e. 
\begin{equation}
\label{eq:NOE_fix_exact} 
\begin{split}
\frac{1}{2}\left(1+\epsilon^{\infty}_{NOE},1-\epsilon^{\infty}_{NOE}\right)\otimes\frac{1}{2}\left(1,1\right)\\
\xrightarrow{\Gamma_2} \frac{1}{4}\left(2p_2,1+\epsilon^{\infty}_{NOE},1-\epsilon^{\infty}_{NOE},2(1-p_2)\right)\\
\xrightarrow{CMS} \frac{1}{2}\left(\frac{1+2p_2+\epsilon^{\infty}_{NOE}}{2},\frac{3-2p_2-\epsilon^{\infty}_{NOE}}{2}\right)\otimes \frac{1}{2}\left(1,1\right).
\end{split}
\end{equation}
The assymptotic polarization should hence obey 
\begin{eqnarray}
2p_2&=&\frac{1+2p_2+\epsilon^{\infty}_{NOE}}{2}\nonumber\\
&\Rightarrow&\epsilon^\infty_{NOE}=2p_2-1=\tanh2\xi
\end{eqnarray}

\section{Appendix C: NOE with multiple reset qubits}

We generalize our description of NOE to a system of $n$ qubits. The description consists of the iteration of two steps. In the first step, we saturate (totally mix) all the qubits with the exception of the first qubit. In the second step, we apply the state-reset $\Gamma_n$. Under this process, the polarization evolution of the first qubit will increase asymptotically to a maximum value. 
To find the cooling limit, which corresponds to the fixed point of the algorithm, we assume that it has a final polarization $\epsilon_{f}$, and we use the condition that if we reapply the two steps of the algorithm that polarization will stay the same. Then, by applying the two steps we have the following:
after the saturation, the diagonal of the state of the system is
\begin{equation}
\label{eq:NOE_g_sat} 
\begin{split}
\xrightarrow{CMS} \frac{1}{2}\left(1+\epsilon_{f},1-\epsilon_{f}\right)\otimes \left[\frac{1}{2}\left(1,1\right)\right]^{\otimes \left(n-1\right)}\\
= \frac{1}{2^n}\left(1+\epsilon_{f}, 1+\epsilon_{f},...,1-\epsilon_{f},1-\epsilon_{f}\right),
\end{split}
\end{equation}
this is a vector with the first $2^{n-1}$ elements equal to $\frac{1}{2^n}\left(1+\epsilon_{f}\right)$ and the last $2^{n-1}$ elements equal to $\frac{1}{2^n}\left(1-\epsilon_{f}\right)$. Then, under the operation $\Gamma_n$, for low polarization, the system will evolve to
\begin{equation}
\label{eq:NOE_g_reset} 
\frac{1}{2^n}\left(1+n\epsilon_{b}, 1+\epsilon_{f},..., 1-\epsilon_{f},1-n\epsilon_{b}\right),
\end{equation}
changing the first and last element to $\frac{1}{2^n}\left(1+n\epsilon_{b}\right)$ and $\frac{1}{2^n}\left(1-n\epsilon_{f}\right)$, respectively.
This results in a polarization $\left[n\epsilon_b+\left(2^{n-1}-1\right)\epsilon_f\right]/2^{n-1}$, which should be equal to the final polarization $\epsilon_f$, thus $\epsilon_f=n\epsilon_b$. This generalized NOE, taken on its own, does not always give better results than the PPA (see section IV-C, where we present the SR$\Gamma_n$-HBAC method: a different way to exploit $\Gamma_n$
to increase the polarization beyond the PPA class of algorithms, in a smaller number of iterations).


\section{Appendix D: NOE-based HBAC}

In this appendix, we present a more practical algorithm based on regular NOE. In this case, the algorithm is limited to use only $\Gamma_2$ to implement regular NOE within a subroutine, in addition to entropy compressions, and qubit-resets. Note that this method is less general than our SR$\Gamma_n$-HBAC, presented in this paper, but still gets better polarization than the PPA.

\subsection{i. The two-qubit case}
When NOE is complemented with a final step
of qubit-reset on the non-target qubit,
the entire system will be cooled (the target qubit will increase its polarization to $\rm{tanh}2\xi$, and the second qubit will be returned to the equilibrium after being saturated).
We name this simple algorithm ``2-NOE-based
HBAC'', and it will be used as a subroutine in this appendix, when cooling
a string of more qubits.

The obtained probabilities to be in state $\left|0\right\rangle$ for each qubit, after applying the ``2-NOE-based
HBAC'', are 
$(1+2\epsilon_b)/2$, and $(1+\epsilon_b)/2$, respectively, in the low polarization case. 
We can denote this probabilities in a more simply way, using
the shifted-and-scaled diagonal terms
$\{2,1\}$ in units of $\epsilon_b$.
 
\subsection{ii. The 3 qubit case}

Here, we show how to use the subroutine ``2-NOE-based
HBAC'' in the three-qubit case, to get probabilities $\{3,2,1\}$ written in the shifted-and-scaled diagonal form in units of $\epsilon_b$, for low polarization.

Let's start from thermal equilibrium, i.e. with $\{1,1,1\}$ in the shifted-and-scaled diagonal form. First, we apply the subroutine ``2-NOE-based
HBAC'' on the second and third qubits, to obtain $\{1,2,1\}$. Then, we cool the target qubit to $2$
using a SORT step (known as 3-bit-compression in the three-qubit case), to get $\{2,1,1\}$.
Applying again a subroutine ``2-NOE-based
HBAC'', we obtain
$\{2,2,1\}$). We can repeat these steps, to achieve 
$\{2.5,2,1\}$). In the same way, another repetition  
yields 
$\{2.75,2,1\}$).  

The polarization enhancement of the target qubit grows asymptotically to a fixed point, corresponding to
polarization $\epsilon^{\infty}$. 
%
After one iteration, in the cooling limit, $\epsilon^{\infty}\to \left(3\epsilon_b+\epsilon^{\infty}\right)/2$
implies that $\epsilon^{\infty}=3\epsilon_b$, yielding the 
final string polarization
$\{3,2,1\}$.

\subsection{iii. The $n$-qubit case}

Using the same process as above, in combination with 3-bit-compressions,
it is easy to obtain a Fibonacci-like series
$\{..., 13,8,5,3,2,1\}$, for low polarization; 
note the only advantage over the SMW-Fibonacci is that the above is better than 
$\{..., 8,5,3,2,1,1\}$ with the same number of qubits. 

Moreover, by using entropy compressions, SORT, rather than 3-bit-compression, as in the original PPA, we obtain polarizations $\{... 24,12,6,3,2,1\}$, improving over the original PPA, namely $\{... 16,8,4,2,1,1\}$.

\bibliographystyle{apsrev4-1}
\bibliography{noe-5}

\begin{thebibliography}{36}%
\makeatletter
\providecommand \@ifxundefined [1]{%
 \@ifx{#1\undefined}
}%
\providecommand \@ifnum [1]{%
 \ifnum #1\expandafter \@firstoftwo
 \else \expandafter \@secondoftwo
 \fi
}%
\providecommand \@ifx [1]{%
 \ifx #1\expandafter \@firstoftwo
 \else \expandafter \@secondoftwo
 \fi
}%
\providecommand \natexlab [1]{#1}%
\providecommand \enquote  [1]{``#1''}%
\providecommand \bibnamefont  [1]{#1}%
\providecommand \bibfnamefont [1]{#1}%
\providecommand \citenamefont [1]{#1}%
\providecommand \href@noop [0]{\@secondoftwo}%
\providecommand \href [0]{\begingroup \@sanitize@url \@href}%
\providecommand \@href[1]{\@@startlink{#1}\@@href}%
\providecommand \@@href[1]{\endgroup#1\@@endlink}%
\providecommand \@sanitize@url [0]{\catcode `\\12\catcode `\$12\catcode
  `\&12\catcode `\#12\catcode `\^12\catcode `\_12\catcode `\%12\relax}%
\providecommand \@@startlink[1]{}%
\providecommand \@@endlink[0]{}%
\providecommand \url  [0]{\begingroup\@sanitize@url \@url }%
\providecommand \@url [1]{\endgroup\@href {#1}{\urlprefix }}%
\providecommand \urlprefix  [0]{URL }%
\providecommand \Eprint [0]{\href }%
\providecommand \doibase [0]{http://dx.doi.org/}%
\providecommand \selectlanguage [0]{\@gobble}%
\providecommand \bibinfo  [0]{\@secondoftwo}%
\providecommand \bibfield  [0]{\@secondoftwo}%
\providecommand \translation [1]{[#1]}%
\providecommand \BibitemOpen [0]{}%
\providecommand \bibitemStop [0]{}%
\providecommand \bibitemNoStop [0]{.\EOS\space}%
\providecommand \EOS [0]{\spacefactor3000\relax}%
\providecommand \BibitemShut  [1]{\csname bibitem#1\endcsname}%
\let\auto@bib@innerbib\@empty
\bibitem [{\citenamefont {Schulman}\ and\ \citenamefont
  {Vazirani}(1999)}]{schulman1999molecular}%
  \BibitemOpen
  \bibfield  {author} {\bibinfo {author} {\bibfnamefont {L.~J.}\ \bibnamefont
  {Schulman}}\ and\ \bibinfo {author} {\bibfnamefont {U.~V.}\ \bibnamefont
  {Vazirani}},\ }in\ \href@noop {} {\emph {\bibinfo {booktitle} {Proceedings of
  the 31th annual ACM symposium on Theory of computing (STOC)}}}\ (\bibinfo
  {organization} {ACM},\ \bibinfo {year} {1999})\ pp.\ \bibinfo {pages}
  {322--329}\BibitemShut {NoStop}%
\bibitem [{\citenamefont {Boykin}\ \emph {et~al.}(2002)\citenamefont {Boykin},
  \citenamefont {Mor}, \citenamefont {Roychowdhury}, \citenamefont {Vatan},\
  and\ \citenamefont {Vrijen}}]{boykin2002algorithmic}%
  \BibitemOpen
  \bibfield  {author} {\bibinfo {author} {\bibfnamefont {P.~O.}\ \bibnamefont
  {Boykin}}, \bibinfo {author} {\bibfnamefont {T.}~\bibnamefont {Mor}},
  \bibinfo {author} {\bibfnamefont {V.}~\bibnamefont {Roychowdhury}}, \bibinfo
  {author} {\bibfnamefont {F.}~\bibnamefont {Vatan}}, \ and\ \bibinfo {author}
  {\bibfnamefont {R.}~\bibnamefont {Vrijen}},\ }\href@noop {} {\bibfield
  {journal} {\bibinfo  {journal} {Proceedings of the National Academy of
  Sciences}\ }\textbf {\bibinfo {volume} {99}},\ \bibinfo {pages} {3388}
  (\bibinfo {year} {2002})}\BibitemShut {NoStop}%
\bibitem [{\citenamefont {Fernandez}\ \emph {et~al.}(2004)\citenamefont
  {Fernandez}, \citenamefont {Lloyd}, \citenamefont {Mor},\ and\ \citenamefont
  {Roychowdhury}}]{fernandez2004algorithmic}%
  \BibitemOpen
  \bibfield  {author} {\bibinfo {author} {\bibfnamefont {J.~M.}\ \bibnamefont
  {Fernandez}}, \bibinfo {author} {\bibfnamefont {S.}~\bibnamefont {Lloyd}},
  \bibinfo {author} {\bibfnamefont {T.}~\bibnamefont {Mor}}, \ and\ \bibinfo
  {author} {\bibfnamefont {V.}~\bibnamefont {Roychowdhury}},\ }\href@noop {}
  {\bibfield  {journal} {\bibinfo  {journal} {International Journal of Quantum
  Information}\ }\textbf {\bibinfo {volume} {2}},\ \bibinfo {pages} {461}
  (\bibinfo {year} {2004})}\BibitemShut {NoStop}%
\bibitem [{\citenamefont {Schulman}\ \emph {et~al.}(2005)\citenamefont
  {Schulman}, \citenamefont {Mor},\ and\ \citenamefont
  {Weinstein}}]{schulman2005physical}%
  \BibitemOpen
  \bibfield  {author} {\bibinfo {author} {\bibfnamefont {L.~J.}\ \bibnamefont
  {Schulman}}, \bibinfo {author} {\bibfnamefont {T.}~\bibnamefont {Mor}}, \
  and\ \bibinfo {author} {\bibfnamefont {Y.}~\bibnamefont {Weinstein}},\
  }\href@noop {} {\bibfield  {journal} {\bibinfo  {journal} {Physical review
  letters}\ }\textbf {\bibinfo {volume} {94}},\ \bibinfo {pages} {120501}
  (\bibinfo {year} {2005})}\BibitemShut {NoStop}%
\bibitem [{\citenamefont {Schulman}\ \emph {et~al.}(2007)\citenamefont
  {Schulman}, \citenamefont {Mor},\ and\ \citenamefont
  {Weinstein}}]{schulman2007physical}%
  \BibitemOpen
  \bibfield  {author} {\bibinfo {author} {\bibfnamefont {L.~J.}\ \bibnamefont
  {Schulman}}, \bibinfo {author} {\bibfnamefont {T.}~\bibnamefont {Mor}}, \
  and\ \bibinfo {author} {\bibfnamefont {Y.}~\bibnamefont {Weinstein}},\
  }\href@noop {} {\bibfield  {journal} {\bibinfo  {journal} {SIAM Journal on
  Computing}\ }\textbf {\bibinfo {volume} {36}},\ \bibinfo {pages} {1729}
  (\bibinfo {year} {2007})}\BibitemShut {NoStop}%
\bibitem [{\citenamefont {Chang}\ \emph {et~al.}(2001)\citenamefont {Chang},
  \citenamefont {Vandersypen},\ and\ \citenamefont {Steffen}}]{chang2001nmr}%
  \BibitemOpen
  \bibfield  {author} {\bibinfo {author} {\bibfnamefont {D.}~\bibnamefont
  {Chang}}, \bibinfo {author} {\bibfnamefont {L.}~\bibnamefont {Vandersypen}},
  \ and\ \bibinfo {author} {\bibfnamefont {M.}~\bibnamefont {Steffen}},\
  }\href@noop {} {\bibfield  {journal} {\bibinfo  {journal} {Chemical physics
  letters}\ }\textbf {\bibinfo {volume} {338}},\ \bibinfo {pages} {337}
  (\bibinfo {year} {2001})}\BibitemShut {NoStop}%
\bibitem [{\citenamefont {Fernandez}\ \emph {et~al.}(2005)\citenamefont
  {Fernandez}, \citenamefont {Mor},\ and\ \citenamefont
  {Weinstein}}]{fernandez2005paramagnetic}%
  \BibitemOpen
  \bibfield  {author} {\bibinfo {author} {\bibfnamefont {J.~M.}\ \bibnamefont
  {Fernandez}}, \bibinfo {author} {\bibfnamefont {T.}~\bibnamefont {Mor}}, \
  and\ \bibinfo {author} {\bibfnamefont {Y.}~\bibnamefont {Weinstein}},\
  }\href@noop {} {\bibfield  {journal} {\bibinfo  {journal} {International
  Journal of Quantum Information}\ }\textbf {\bibinfo {volume} {3}},\ \bibinfo
  {pages} {281} (\bibinfo {year} {2005})}\BibitemShut {NoStop}%
\bibitem [{\citenamefont {Elias}\ \emph
  {et~al.}(2011{\natexlab{a}})\citenamefont {Elias}, \citenamefont {Gilboa},
  \citenamefont {Mor},\ and\ \citenamefont {Weinstein}}]{elias2011heat}%
  \BibitemOpen
  \bibfield  {author} {\bibinfo {author} {\bibfnamefont {Y.}~\bibnamefont
  {Elias}}, \bibinfo {author} {\bibfnamefont {H.}~\bibnamefont {Gilboa}},
  \bibinfo {author} {\bibfnamefont {T.}~\bibnamefont {Mor}}, \ and\ \bibinfo
  {author} {\bibfnamefont {Y.}~\bibnamefont {Weinstein}},\ }\href@noop {}
  {\bibfield  {journal} {\bibinfo  {journal} {Chemical Physics Letters}\
  }\textbf {\bibinfo {volume} {517}},\ \bibinfo {pages} {126} (\bibinfo {year}
  {2011}{\natexlab{a}})}\BibitemShut {NoStop}%
\bibitem [{\citenamefont {Brassard}\ \emph {et~al.}(1156)\citenamefont
  {Brassard}, \citenamefont {Elias}, \citenamefont {Mor},\ and\ \citenamefont
  {Weinstein}}]{brassard2014prospects}%
  \BibitemOpen
  \bibfield  {author} {\bibinfo {author} {\bibfnamefont {G.}~\bibnamefont
  {Brassard}}, \bibinfo {author} {\bibfnamefont {Y.}~\bibnamefont {Elias}},
  \bibinfo {author} {\bibfnamefont {T.}~\bibnamefont {Mor}}, \ and\ \bibinfo
  {author} {\bibfnamefont {Y.}~\bibnamefont {Weinstein}},\ }\href@noop {}
  {\bibfield  {journal} {\bibinfo  {journal} {The European Physical Journal
  Plus}\ }\textbf {\bibinfo {volume} {129}},\ \bibinfo {pages} {1} (\bibinfo
  {year} {2014 and arXiv:quant-ph/0511156})}\BibitemShut {NoStop}%
\bibitem [{\citenamefont {Baugh}\ \emph {et~al.}(2005)\citenamefont {Baugh},
  \citenamefont {Moussa}, \citenamefont {Ryan}, \citenamefont {Nayak},\ and\
  \citenamefont {Laflamme}}]{baugh2005experimental}%
  \BibitemOpen
  \bibfield  {author} {\bibinfo {author} {\bibfnamefont {J.}~\bibnamefont
  {Baugh}}, \bibinfo {author} {\bibfnamefont {O.}~\bibnamefont {Moussa}},
  \bibinfo {author} {\bibfnamefont {C.~A.}\ \bibnamefont {Ryan}}, \bibinfo
  {author} {\bibfnamefont {A.}~\bibnamefont {Nayak}}, \ and\ \bibinfo {author}
  {\bibfnamefont {R.}~\bibnamefont {Laflamme}},\ }\href@noop {} {\bibfield
  {journal} {\bibinfo  {journal} {Nature}\ }\textbf {\bibinfo {volume} {438}},\
  \bibinfo {pages} {470} (\bibinfo {year} {2005})}\BibitemShut {NoStop}%
\bibitem [{\citenamefont {Ryan}\ \emph {et~al.}(2008)\citenamefont {Ryan},
  \citenamefont {Moussa}, \citenamefont {Baugh},\ and\ \citenamefont
  {Laflamme}}]{Ryan:2008qf}%
  \BibitemOpen
  \bibfield  {author} {\bibinfo {author} {\bibfnamefont {C.~A.}\ \bibnamefont
  {Ryan}}, \bibinfo {author} {\bibfnamefont {O.}~\bibnamefont {Moussa}},
  \bibinfo {author} {\bibfnamefont {J.}~\bibnamefont {Baugh}}, \ and\ \bibinfo
  {author} {\bibfnamefont {R.}~\bibnamefont {Laflamme}},\ }\href@noop {}
  {\bibfield  {journal} {\bibinfo  {journal} {Physical review letters}\
  }\textbf {\bibinfo {volume} {100}},\ \bibinfo {pages} {140501} (\bibinfo
  {year} {2008})}\BibitemShut {NoStop}%
\bibitem [{\citenamefont {Park}\ \emph {et~al.}(2015)\citenamefont {Park},
  \citenamefont {Feng}, \citenamefont {Rahimi}, \citenamefont {Labruyere},
  \citenamefont {Shibata}, \citenamefont {Nakazawa}, \citenamefont {Sato},
  \citenamefont {Takui}, \citenamefont {Laflamme},\ and\ \citenamefont
  {Baugh}}]{park2014hyperfine}%
  \BibitemOpen
  \bibfield  {author} {\bibinfo {author} {\bibfnamefont {D.~K.}\ \bibnamefont
  {Park}}, \bibinfo {author} {\bibfnamefont {G.}~\bibnamefont {Feng}}, \bibinfo
  {author} {\bibfnamefont {R.}~\bibnamefont {Rahimi}}, \bibinfo {author}
  {\bibfnamefont {S.}~\bibnamefont {Labruyere}}, \bibinfo {author}
  {\bibfnamefont {T.}~\bibnamefont {Shibata}}, \bibinfo {author} {\bibfnamefont
  {S.}~\bibnamefont {Nakazawa}}, \bibinfo {author} {\bibfnamefont
  {K.}~\bibnamefont {Sato}}, \bibinfo {author} {\bibfnamefont {T.}~\bibnamefont
  {Takui}}, \bibinfo {author} {\bibfnamefont {R.}~\bibnamefont {Laflamme}}, \
  and\ \bibinfo {author} {\bibfnamefont {J.}~\bibnamefont {Baugh}},\
  }\href@noop {} {\bibfield  {journal} {\bibinfo  {journal} {arXiv:1501.00082}\
  } (\bibinfo {year} {2015})}\BibitemShut {NoStop}%
\bibitem [{\citenamefont {Park}\ \emph {et~al.}(2016)\citenamefont {Park},
  \citenamefont {Rodriguez-Briones}, \citenamefont {Feng}, \citenamefont
  {Rahimi}, \citenamefont {Baugh},\ and\ \citenamefont
  {Laflamme}}]{park2015heat}%
  \BibitemOpen
  \bibfield  {author} {\bibinfo {author} {\bibfnamefont {D.~K.}\ \bibnamefont
  {Park}}, \bibinfo {author} {\bibfnamefont {N.~A.}\ \bibnamefont
  {Rodriguez-Briones}}, \bibinfo {author} {\bibfnamefont {G.}~\bibnamefont
  {Feng}}, \bibinfo {author} {\bibfnamefont {R.}~\bibnamefont {Rahimi}},
  \bibinfo {author} {\bibfnamefont {J.}~\bibnamefont {Baugh}}, \ and\ \bibinfo
  {author} {\bibfnamefont {R.}~\bibnamefont {Laflamme}},\ }in\ \href@noop {}
  {\emph {\bibinfo {booktitle} {Electron Spin Resonance (ESR) Based Quantum
  Computing}}}\ (\bibinfo  {publisher} {Springer},\ \bibinfo {year} {2016})\
  pp.\ \bibinfo {pages} {227--255}\BibitemShut {NoStop}%
\bibitem [{\citenamefont {Mor}\ \emph {et~al.}(2005)\citenamefont {Mor},
  \citenamefont {Fernandez}, \citenamefont {Lloyd}, \citenamefont {Mor},
  \citenamefont {Roychowdhury},\ and\ \citenamefont
  {Weinstein}}]{mor2005patent}%
  \BibitemOpen
  \bibfield  {author} {\bibinfo {author} {\bibfnamefont {T.}~\bibnamefont
  {Mor}}, \bibinfo {author} {\bibfnamefont {J.~M.}\ \bibnamefont {Fernandez}},
  \bibinfo {author} {\bibfnamefont {S.}~\bibnamefont {Lloyd}}, \bibinfo
  {author} {\bibfnamefont {T.}~\bibnamefont {Mor}}, \bibinfo {author}
  {\bibfnamefont {V.}~\bibnamefont {Roychowdhury}}, \ and\ \bibinfo {author}
  {\bibfnamefont {Y.}~\bibnamefont {Weinstein}},\ }\href@noop {} {\bibfield
  {journal} {\bibinfo  {journal} {USA PATENT}\ } (\bibinfo {year}
  {2005})}\BibitemShut {NoStop}%
\bibitem [{\citenamefont {S{\o}rensen}(1991)}]{sorensen1991entropy}%
  \BibitemOpen
  \bibfield  {author} {\bibinfo {author} {\bibfnamefont {O.~W.}\ \bibnamefont
  {S{\o}rensen}},\ }\href@noop {} {\bibfield  {journal} {\bibinfo  {journal}
  {Journal of Magnetic Resonance (1969)}\ }\textbf {\bibinfo {volume} {93}},\
  \bibinfo {pages} {648} (\bibinfo {year} {1991})}\BibitemShut {NoStop}%
\bibitem [{\citenamefont {Peres}(1992)}]{peres1992_590}%
  \BibitemOpen
  \bibfield  {author} {\bibinfo {author} {\bibfnamefont {Y.}~\bibnamefont
  {Peres}},\ }\href@noop {} {\bibfield  {journal} {\bibinfo  {journal} {The
  Annals of Statistics}\ }\textbf {\bibinfo {volume} {20}},\ \bibinfo {pages}
  {590} (\bibinfo {year} {1992})}\BibitemShut {NoStop}%
\bibitem [{\citenamefont {Von~Neumann}(1951)}]{von195113}%
  \BibitemOpen
  \bibfield  {author} {\bibinfo {author} {\bibfnamefont {J.}~\bibnamefont
  {Von~Neumann}},\ }\href@noop {} {\bibfield  {journal} {\bibinfo  {journal}
  {National Bureau of Standards Applied Mathematics Series}\ ,\ \bibinfo
  {pages} {12:36}} (\bibinfo {year} {1951})}\BibitemShut {NoStop}%
\bibitem [{\citenamefont {Moussa}(2005)}]{moussa:2005}%
  \BibitemOpen
  \bibfield  {author} {\bibinfo {author} {\bibfnamefont {O.}~\bibnamefont
  {Moussa}},\ }\emph {\bibinfo {title} {On heat-bath algorithmic cooling and
  its implementation in solid-state NMR}},\ \href@noop {} {\bibinfo {type}
  {Master of science in physics thesis}},\ \bibinfo  {school} {University of
  Waterloo} (\bibinfo {year} {2005})\BibitemShut {NoStop}%
\bibitem [{\citenamefont {Elias}\ \emph {et~al.}(2006)\citenamefont {Elias},
  \citenamefont {Fernandez}, \citenamefont {Mor},\ and\ \citenamefont
  {Weinstein}}]{elias2006optimal}%
  \BibitemOpen
  \bibfield  {author} {\bibinfo {author} {\bibfnamefont {Y.}~\bibnamefont
  {Elias}}, \bibinfo {author} {\bibfnamefont {J.~M.}\ \bibnamefont
  {Fernandez}}, \bibinfo {author} {\bibfnamefont {T.}~\bibnamefont {Mor}}, \
  and\ \bibinfo {author} {\bibfnamefont {Y.}~\bibnamefont {Weinstein}},\
  }\href@noop {} {\bibfield  {journal} {\bibinfo  {journal} {Israel Journal of
  Chemistry}\ }\textbf {\bibinfo {volume} {46}},\ \bibinfo {pages} {371}
  (\bibinfo {year} {2006})}\BibitemShut {NoStop}%
\bibitem [{\citenamefont {Elias}\ \emph
  {et~al.}(2011{\natexlab{b}})\citenamefont {Elias}, \citenamefont {Mor},\ and\
  \citenamefont {Weinstein}}]{elias2011semioptimal}%
  \BibitemOpen
  \bibfield  {author} {\bibinfo {author} {\bibfnamefont {Y.}~\bibnamefont
  {Elias}}, \bibinfo {author} {\bibfnamefont {T.}~\bibnamefont {Mor}}, \ and\
  \bibinfo {author} {\bibfnamefont {Y.}~\bibnamefont {Weinstein}},\ }\href@noop
  {} {\bibfield  {journal} {\bibinfo  {journal} {Physical Review A}\ }\textbf
  {\bibinfo {volume} {83}},\ \bibinfo {pages} {042340} (\bibinfo {year}
  {2011}{\natexlab{b}})}\BibitemShut {NoStop}%
\bibitem [{\citenamefont {Kaye}(2007)}]{kaye:2007}%
  \BibitemOpen
  \bibfield  {author} {\bibinfo {author} {\bibfnamefont {P.}~\bibnamefont
  {Kaye}},\ }\href@noop {} {\bibfield  {journal} {\bibinfo  {journal} {Quantum
  Information Processing}\ }\textbf {\bibinfo {volume} {6}} (\bibinfo {year}
  {2007})}\BibitemShut {NoStop}%
\bibitem [{\citenamefont {Frey}\ \emph {et~al.}(2014)\citenamefont {Frey},
  \citenamefont {Funo},\ and\ \citenamefont {Hotta}}]{frey2014strong}%
  \BibitemOpen
  \bibfield  {author} {\bibinfo {author} {\bibfnamefont {M.}~\bibnamefont
  {Frey}}, \bibinfo {author} {\bibfnamefont {K.}~\bibnamefont {Funo}}, \ and\
  \bibinfo {author} {\bibfnamefont {M.}~\bibnamefont {Hotta}},\ }\href@noop {}
  {\bibfield  {journal} {\bibinfo  {journal} {Physical Review E}\ }\textbf
  {\bibinfo {volume} {90}},\ \bibinfo {pages} {012127} (\bibinfo {year}
  {2014})}\BibitemShut {NoStop}%
\bibitem [{\citenamefont {Brunner}\ \emph {et~al.}(2014)\citenamefont
  {Brunner}, \citenamefont {Huber}, \citenamefont {Linden}, \citenamefont
  {Popescu}, \citenamefont {Silva},\ and\ \citenamefont
  {Skrzypczyk}}]{brunner2014entanglement}%
  \BibitemOpen
  \bibfield  {author} {\bibinfo {author} {\bibfnamefont {N.}~\bibnamefont
  {Brunner}}, \bibinfo {author} {\bibfnamefont {M.}~\bibnamefont {Huber}},
  \bibinfo {author} {\bibfnamefont {N.}~\bibnamefont {Linden}}, \bibinfo
  {author} {\bibfnamefont {S.}~\bibnamefont {Popescu}}, \bibinfo {author}
  {\bibfnamefont {R.}~\bibnamefont {Silva}}, \ and\ \bibinfo {author}
  {\bibfnamefont {P.}~\bibnamefont {Skrzypczyk}},\ }\href@noop {} {\bibfield
  {journal} {\bibinfo  {journal} {Physical Review E}\ }\textbf {\bibinfo
  {volume} {89}},\ \bibinfo {pages} {032115} (\bibinfo {year}
  {2014})}\BibitemShut {NoStop}%
\bibitem [{\citenamefont {Perarnau-Llobet}\ \emph {et~al.}(2015)\citenamefont
  {Perarnau-Llobet}, \citenamefont {Hovhannisyan}, \citenamefont {Huber},
  \citenamefont {Skrzypczyk}, \citenamefont {Brunner},\ and\ \citenamefont
  {Ac{\'\i}n}}]{perarnau2015extractable}%
  \BibitemOpen
  \bibfield  {author} {\bibinfo {author} {\bibfnamefont {M.}~\bibnamefont
  {Perarnau-Llobet}}, \bibinfo {author} {\bibfnamefont {K.~V.}\ \bibnamefont
  {Hovhannisyan}}, \bibinfo {author} {\bibfnamefont {M.}~\bibnamefont {Huber}},
  \bibinfo {author} {\bibfnamefont {P.}~\bibnamefont {Skrzypczyk}}, \bibinfo
  {author} {\bibfnamefont {N.}~\bibnamefont {Brunner}}, \ and\ \bibinfo
  {author} {\bibfnamefont {A.}~\bibnamefont {Ac{\'\i}n}},\ }\href@noop {}
  {\bibfield  {journal} {\bibinfo  {journal} {Physical Review X}\ }\textbf
  {\bibinfo {volume} {5}},\ \bibinfo {pages} {041011} (\bibinfo {year}
  {2015})}\BibitemShut {NoStop}%
\bibitem [{\citenamefont {Rodriguez-Briones}\ \emph {et~al.}(2015)\citenamefont
  {Rodriguez-Briones}, \citenamefont {Li}, \citenamefont {Peng}, \citenamefont
  {Mor}, \citenamefont {Weinstein},\ and\ \citenamefont
  {Laflamme}}]{rodriguez2015comments}%
  \BibitemOpen
  \bibfield  {author} {\bibinfo {author} {\bibfnamefont {N.~A.}\ \bibnamefont
  {Rodriguez-Briones}}, \bibinfo {author} {\bibfnamefont {J.}~\bibnamefont
  {Li}}, \bibinfo {author} {\bibfnamefont {X.}~\bibnamefont {Peng}}, \bibinfo
  {author} {\bibfnamefont {T.}~\bibnamefont {Mor}}, \bibinfo {author}
  {\bibfnamefont {Y.}~\bibnamefont {Weinstein}}, \ and\ \bibinfo {author}
  {\bibfnamefont {R.}~\bibnamefont {Laflamme}},\ }\href@noop {} {\bibfield
  {journal} {\bibinfo  {journal} {arXiv preprint arXiv:1506.01778}\ } (\bibinfo
  {year} {2015})}\BibitemShut {NoStop}%
\bibitem [{\citenamefont {Liuzzo-Scorpo}\ \emph {et~al.}(2016)\citenamefont
  {Liuzzo-Scorpo}, \citenamefont {Correa}, \citenamefont {Schmidt},\ and\
  \citenamefont {Adesso}}]{liuzzo2016thermodynamics}%
  \BibitemOpen
  \bibfield  {author} {\bibinfo {author} {\bibfnamefont {P.}~\bibnamefont
  {Liuzzo-Scorpo}}, \bibinfo {author} {\bibfnamefont {L.~A.}\ \bibnamefont
  {Correa}}, \bibinfo {author} {\bibfnamefont {R.}~\bibnamefont {Schmidt}}, \
  and\ \bibinfo {author} {\bibfnamefont {G.}~\bibnamefont {Adesso}},\
  }\href@noop {} {\bibfield  {journal} {\bibinfo  {journal} {Entropy}\ }\textbf
  {\bibinfo {volume} {18}},\ \bibinfo {pages} {48} (\bibinfo {year}
  {2016})}\BibitemShut {NoStop}%
\bibitem [{\citenamefont {Friis}\ \emph {et~al.}(2016)\citenamefont {Friis},
  \citenamefont {Huber},\ and\ \citenamefont
  {Perarnau-Llobet}}]{PhysRevE.93.042135}%
  \BibitemOpen
  \bibfield  {author} {\bibinfo {author} {\bibfnamefont {N.}~\bibnamefont
  {Friis}}, \bibinfo {author} {\bibfnamefont {M.}~\bibnamefont {Huber}}, \ and\
  \bibinfo {author} {\bibfnamefont {M.}~\bibnamefont {Perarnau-Llobet}},\
  }\href {\doibase 10.1103/PhysRevE.93.042135} {\bibfield  {journal} {\bibinfo
  {journal} {Phys. Rev. E}\ }\textbf {\bibinfo {volume} {93}},\ \bibinfo
  {pages} {042135} (\bibinfo {year} {2016})}\BibitemShut {NoStop}%
\bibitem [{\citenamefont {Goold}\ \emph {et~al.}(2016)\citenamefont {Goold},
  \citenamefont {Huber}, \citenamefont {Riera}, \citenamefont {del Rio},\ and\
  \citenamefont {Skrzypczyk}}]{goold2016role}%
  \BibitemOpen
  \bibfield  {author} {\bibinfo {author} {\bibfnamefont {J.}~\bibnamefont
  {Goold}}, \bibinfo {author} {\bibfnamefont {M.}~\bibnamefont {Huber}},
  \bibinfo {author} {\bibfnamefont {A.}~\bibnamefont {Riera}}, \bibinfo
  {author} {\bibfnamefont {L.}~\bibnamefont {del Rio}}, \ and\ \bibinfo
  {author} {\bibfnamefont {P.}~\bibnamefont {Skrzypczyk}},\ }\href@noop {}
  {\bibfield  {journal} {\bibinfo  {journal} {Journal of Physics A:
  Mathematical and Theoretical}\ }\textbf {\bibinfo {volume} {49}},\ \bibinfo
  {pages} {143001} (\bibinfo {year} {2016})}\BibitemShut {NoStop}%
\bibitem [{\citenamefont {Dillenschneider}\ and\ \citenamefont
  {Lutz}(2009)}]{dillenschneider2009energetics}%
  \BibitemOpen
  \bibfield  {author} {\bibinfo {author} {\bibfnamefont {R.}~\bibnamefont
  {Dillenschneider}}\ and\ \bibinfo {author} {\bibfnamefont {E.}~\bibnamefont
  {Lutz}},\ }\href@noop {} {\bibfield  {journal} {\bibinfo  {journal} {EPL
  (Europhysics Letters)}\ }\textbf {\bibinfo {volume} {88}},\ \bibinfo {pages}
  {50003} (\bibinfo {year} {2009})}\BibitemShut {NoStop}%
\bibitem [{\citenamefont {Overhauser}(1953)}]{Overhauser:1953fk}%
  \BibitemOpen
  \bibfield  {author} {\bibinfo {author} {\bibfnamefont {A.~W.}\ \bibnamefont
  {Overhauser}},\ }\href@noop {} {\bibfield  {journal} {\bibinfo  {journal}
  {Physical Review}\ }\textbf {\bibinfo {volume} {89}},\ \bibinfo {pages} {689}
  (\bibinfo {year} {1953})}\BibitemShut {NoStop}%
\bibitem [{\citenamefont {Rodr\'{\i}guez-Briones}\ and\ \citenamefont
  {Laflamme}(2016)}]{PhysRevLett.116.170501}%
  \BibitemOpen
  \bibfield  {author} {\bibinfo {author} {\bibfnamefont {N.~A.}\ \bibnamefont
  {Rodr\'{\i}guez-Briones}}\ and\ \bibinfo {author} {\bibfnamefont
  {R.}~\bibnamefont {Laflamme}},\ }\href {\doibase
  10.1103/PhysRevLett.116.170501} {\bibfield  {journal} {\bibinfo  {journal}
  {Phys. Rev. Lett.}\ }\textbf {\bibinfo {volume} {116}},\ \bibinfo {pages}
  {170501} (\bibinfo {year} {2016})}\BibitemShut {NoStop}%
\bibitem [{\citenamefont {Raeisi}\ and\ \citenamefont
  {Mosca}(2015)}]{raeisi2015asymptotic}%
  \BibitemOpen
  \bibfield  {author} {\bibinfo {author} {\bibfnamefont {S.}~\bibnamefont
  {Raeisi}}\ and\ \bibinfo {author} {\bibfnamefont {M.}~\bibnamefont {Mosca}},\
  }\href@noop {} {\bibfield  {journal} {\bibinfo  {journal} {Physical Review
  Letters}\ }\textbf {\bibinfo {volume} {114}},\ \bibinfo {pages} {100404}
  (\bibinfo {year} {2015})}\BibitemShut {NoStop}%
\bibitem [{\citenamefont {Li}\ \emph {et~al.}(2016)\citenamefont {Li},
  \citenamefont {Lu}, \citenamefont {Luo}, \citenamefont {Laflamme},
  \citenamefont {Peng},\ and\ \citenamefont {Du}}]{li2014maximally}%
  \BibitemOpen
  \bibfield  {author} {\bibinfo {author} {\bibfnamefont {J.}~\bibnamefont
  {Li}}, \bibinfo {author} {\bibfnamefont {D.}~\bibnamefont {Lu}}, \bibinfo
  {author} {\bibfnamefont {Z.}~\bibnamefont {Luo}}, \bibinfo {author}
  {\bibfnamefont {R.}~\bibnamefont {Laflamme}}, \bibinfo {author}
  {\bibfnamefont {X.}~\bibnamefont {Peng}}, \ and\ \bibinfo {author}
  {\bibfnamefont {J.}~\bibnamefont {Du}},\ }\href {\doibase
  10.1103/PhysRevA.94.012312} {\bibfield  {journal} {\bibinfo  {journal} {Phys.
  Rev. A}\ }\textbf {\bibinfo {volume} {94}},\ \bibinfo {pages} {012312}
  (\bibinfo {year} {2016})}\BibitemShut {NoStop}%
\bibitem [{\citenamefont {Poyatos}\ \emph {et~al.}(1996)\citenamefont
  {Poyatos}, \citenamefont {Cirac},\ and\ \citenamefont
  {Zoller}}]{PhysRevLett.77.4728}%
  \BibitemOpen
  \bibfield  {author} {\bibinfo {author} {\bibfnamefont {J.~F.}\ \bibnamefont
  {Poyatos}}, \bibinfo {author} {\bibfnamefont {J.~I.}\ \bibnamefont {Cirac}},
  \ and\ \bibinfo {author} {\bibfnamefont {P.}~\bibnamefont {Zoller}},\ }\href
  {\doibase 10.1103/PhysRevLett.77.4728} {\bibfield  {journal} {\bibinfo
  {journal} {Phys. Rev. Lett.}\ }\textbf {\bibinfo {volume} {77}},\ \bibinfo
  {pages} {4728} (\bibinfo {year} {1996})}\BibitemShut {NoStop}%
\bibitem [{\citenamefont {Knill}\ \emph {et~al.}(2000)\citenamefont {Knill},
  \citenamefont {Laflamme},\ and\ \citenamefont {Viola}}]{knill2000theory}%
  \BibitemOpen
  \bibfield  {author} {\bibinfo {author} {\bibfnamefont {E.}~\bibnamefont
  {Knill}}, \bibinfo {author} {\bibfnamefont {R.}~\bibnamefont {Laflamme}}, \
  and\ \bibinfo {author} {\bibfnamefont {L.}~\bibnamefont {Viola}},\
  }\href@noop {} {\bibfield  {journal} {\bibinfo  {journal} {Physical Review
  Letters}\ }\textbf {\bibinfo {volume} {84}},\ \bibinfo {pages} {2525}
  (\bibinfo {year} {2000})}\BibitemShut {NoStop}%
\bibitem [{\citenamefont {Solomon}(1955)}]{Solomon:1955uq}%
  \BibitemOpen
  \bibfield  {author} {\bibinfo {author} {\bibfnamefont {I.}~\bibnamefont
  {Solomon}},\ }\href@noop {} {\bibfield  {journal} {\bibinfo  {journal} {Phys.
  Rev.}\ }\textbf {\bibinfo {volume} {99}},\ \bibinfo {pages} {559} (\bibinfo
  {year} {1955})}\BibitemShut {NoStop}%
\end{thebibliography}%

\end{document}